\documentclass[pdflatex,sn-mathphys-num]{sn-jnl}

\usepackage{graphicx}%
\newcommand{\todosara}[1]{\todo[color=purple!40]{\color{black} SARA: \small{{#1}}}}

\usepackage{multirow}%
\usepackage{amsmath,amssymb,amsfonts}%
\usepackage{amsthm}%
\usepackage{mathrsfs}%
\usepackage[title]{appendix}%
\usepackage{xcolor}%
\usepackage{textcomp}%
\usepackage{manyfoot}%
\usepackage{booktabs}%
\usepackage{algorithm}%
\usepackage{algorithmicx}%
\usepackage{algpseudocode}%
\usepackage{listings}%
\usepackage{tcolorbox}
\usepackage{arydshln}
 \usepackage{comment}

\theoremstyle{thmstyleone}%
\newtheorem{theorem}{Theorem}
\theoremstyle{thmstyletwo}%
\newtheorem{example}{Example}%
\newtheorem{remark}{Remark}%

\theoremstyle{thmstylethree}%
\newtheorem{definition}{Definition}%
\usepackage{xfrac}
\usepackage{relsize}

\usepackage[textsize=small]{todonotes}
\usepackage{xfrac}

\newcommand{\Fset}{\mathbb{F}}

\newcommand{\card}[1]{\left|#1\right|}

\usepackage[normalem]{ulem}
\begin{document}

\title[Decoding Algorithms]{Decoding Algorithms for MDS Array Codes}


\author*[1]{\fnm{Sara} \sur{D. Cardell}}\email{sd.cardell@unesp.br}
\equalcont{These authors contributed equally to this work.}

\author[2]{\fnm{Gustavo} \sur{Terra Bastos}}\email{gtbastos@ufsj.edu.br}
\equalcont{These authors contributed equally to this work.}

\author[3]{\fnm{Cintya} \sur{	 Wink de Oliveira Benedito }\email{cintya.benedito@unesp.br }}
\equalcont{These authors contributed equally to this work.}

\affil[1]{\orgdiv{Department of Mathematics}, \orgname{Institute of Geosciences and Exact Sciences, Unesp}, \orgaddress{\street{Av. 24-A nº. 1515 }, \city{Rio Claro}, \postcode{13506-900}, \state{SP}, \country{Brazil}}}

\affil[2]{\orgdiv{Department of Mathematics and Statistics}, \orgname{Federal University of São João del-Rei}, \orgaddress{\street{170 Frei Orlando Square}, \city{São João del-Rei}, \postcode{36307-352}, \state{MG}, \country{Brazil}}}

\affil[3]{\orgdiv{School of Engineering}, \orgname{Unesp}, \orgaddress{\street{505 Profa. Isette Correa Fontão Avenue}, \city{São João da Boa Vista}, \postcode{13876-750}, \state{SP}, \country{Brazil}}}



 \abstract{
 We study decoding procedures for a family of MDS array codes previously constructed from the Kronecker product of a superregular matrix and a non-singular matrix over a finite field. By exploiting the particular structure of their parity-check matrices, we develop decoding algorithms for different channel models. For the erasure channel, we provide an algorithm capable of recovering any pattern of up to $n-k$ symbol erasures. For the $q$-ary symmetric channel, we investigate the decoding of one and two symbol errors and give explicit procedures for determining their locations and values. We also consider the particular case in which the superregular matrix is a Vandermonde matrix, showing how its additional algebraic structure can be exploited in the decoding process. Explicit examples over different finite fields are provided to illustrate the proposed procedures.

 }





\maketitle
  \section{Introduction}


Reliable transmission and storage of information require mechanisms capable of recovering data affected by noise, interference, or failures. Coding techniques address this problem by introducing structured redundancy into the data, enabling the detection and correction of errors at the receiver or during data recovery \cite{huffman,sloane}. The error-correcting capability of a code is closely related to its minimum Hamming distance, which determines the largest number of errors that can be uniquely corrected. In particular, codes whose minimum distance attains the Singleton bound are known as maximum distance separable (MDS) codes and achieve the largest possible minimum distance for a given code length and dimension \cite{huffman,handbook,sloane}.

Array codes provide a natural framework for organizing code symbols into multidimensional structures, facilitating the representation and recovery of encoded data. These features have motivated their use in applications requiring efficient error control and decoding \cite{handbook,Farrell1992}. When array codes also satisfy the MDS property, they combine the flexibility of the array representation with the optimal distance of MDS codes, making them particularly suitable for storage systems that require reliable and efficient data recovery.

MDS array codes have been constructed using different algebraic and combinatorial approaches. Classical contributions include the construction of array codes for phased burst correction by Blaum and Roth \cite{BlaumRoth1993} and the subsequent development of MDS array codes with independent parity symbols in \cite{BlaumBruckVardy1996}. More recent constructions have addressed requirements arising in distributed storage systems, including efficient repair and reduced sub-packetization levels \cite{mds_subpacketization}, as well as constructions based on deleted circulant matrices have been used to obtain MDS array codes and, subsequently, optimal locally repairable array codes \cite{fang}. Among algebraic approaches, constructions based on superregular matrices provide a natural framework for deriving MDS array codes together with their decoding procedures. In this context, MDS array codes whose parity-check matrices are obtained from block-superregular matrices composed of powers of a companion matrix, and also develop a decoding algorithm for the resulting codes \cite{sara}. Following these developments, \cite{cintyaarxiv} considers MDS array codes constructed from superregular matrices, with particular emphasis on Vandermonde matrices, and develops decoding algorithms for correcting symbol errors without prior knowledge of their locations. In addition,  was proposed in \cite{Bastos2025} constructions of superregular and block-superregular matrices based on Kronecker products, providing a general algebraic framework for generating the matrix structures underlying the MDS array codes considered in this work.

The $q$-ary symmetric channel ($q$-SC) and the $q$-ary erasure channel ($q$-EC) are two standard channel models for coding over the finite field $\mathbb{F}_q$. In the $q$-SC, each transmitted symbol is received correctly with probability $1-p$, while with probability $p$ it is replaced by one of the other $q-1$ symbols in $\mathbb{F}_q$, each with probability $p/(q-1)$ \cite{Weidmann2012,Pernice2025}. In the $q$-EC, the output alphabet is $\mathbb{F}_q\cup{\varepsilon}$, and each transmitted symbol is received correctly with probability $1-\rho$ or erased with probability $\rho$, in which case the receiver observes the erasure symbol $\varepsilon$ \cite{Liva2013,Shen2019,Chan2025}. These channel models lead to fundamentally different decoding scenarios: in the $q$-EC, the locations of the erased symbols are known to the decoder, whereas in the $q$-SC, the locations of the corrupted symbols are unknown. Moreover, a code with minimum distance $d$ can uniquely recover any pattern of up to $d-1$ symbol erasures \cite{Liva2013}.

In this work, we exploit the algebraic structure underlying the MDS array codes associated with the construction of superregular and block superregular matrices based on the Kronecker product of a superregular matrix and a non-singular matrix over a finite field \cite{Bastos2025} to derive decoding procedures for these two channel models. For the $q$-EC, we develop an algorithm capable of recovering any pattern of up to $n-k$ symbol erasures. For the $q$-SC, where the error locations are not known to the decoder, we develop explicit procedures for correcting one and two symbol errors by determining both their locations and values. We further investigate the particular case in which the superregular matrix is Vandermonde, exploiting its additional algebraic structure to simplify the decoding procedures. These results highlight how the algebraic structure of the parity-check matrix can be directly exploited to design decoding algorithms for MDS array codes under substantially different channel conditions.





\section{Preliminaries}\label{sec:prel}
Let $\mathbb{F}_q$ be the finite field with $q$ elements.
In this section we introduce the main notions of this work.

\subsection{Linear codes}


We now recall the definition of  $\mathbb{F}_q$-linear code over the alphabet $\mathbb{F}_q^b$ \cite{Louidor2006,Blaum1999}. 

\begin{definition}\label{Prel:def:1}
 Let $b$ be a positive integer.
 A code $\mathcal{C}$ is said to be an $\Fset_q$-linear code of length $n$ over $\Fset_q^b$ if it is a linear subspace of the vector space
$\Fset_q^{nb}$.
Therefore, the codewords of $\mathcal{C}$ can be seen as codewords of length $nb$ over~$\Fset_q$.
\end{definition}

Let us denote by \( \mathcal{C}_{\mathbb{F}_q} \) and \( \mathcal{C}_{\mathbb{F}_q^b} \) the code viewed over the different alphabets, $\mathbb{F}_q$ and $\mathbb{F}_q^b$, respectively.  
Both \( \mathcal{C}_{\mathbb{F}_q} \) and \( \mathcal{C}_{\mathbb{F}_q^b} \) refer to the same set of codewords, but considered over different alphabets.

Let $[N,K,D]$ denote the parameters of the code $\mathcal{C}_{\Fset_q}$.
The number $k = \log_{q^{b}}\card{\mathcal{C}_{\Fset_q^b}}$ is called the normalized dimension (or just dimension) of $\mathcal{C}_{\Fset_q^b}$ \cite{Louidor2006}. 
If $b$ divides $K$ then $k = {K}/{b}$.
Thus, the parameters of the code $\mathcal{C}_{\mathbb{F}_{q}^{b}}$ are $[n,k,d]$ over $\Fset_{q}^b$, where $d$ is the minimum distance and $n={N}/{b}$.
The minimum distance of $\mathcal{C}_{\Fset_q^b}$  
is measured with respect to the symbols of $\Fset_{q}^{b}$ (see \cite{Blaum1999}).



\begin{example}  \label{Prel:ex:1}
Consider   the following matrix
\[
  G
  =
  \left[
    \begin{array}{cc|cc|cc|cc}
      1 & 0 & 0 & 0 & 1 & 1 & 1 & 1 \\
      0 & 1 & 0 & 0 & 1 & 0 & 1 & 0 \\ 
      0 & 0 & 1 & 0 & 1 & 1 & 0 & 1 \\
      0 & 0 & 0 & 1 & 0 & 1 & 1 & 0 \\
    \end{array}
  \right].
\]
This matrix   is a generator matrix of a binary code $\mathcal{C}_{\Fset_2}$ with parameters $[8,4, 3]$.
If we divide the bits of each codeword into groups of two elements, then the codewords, over the new alphabet $\mathbb{F}_2^2$, have length $4$ and  the normalized dimension is $2$. 
The generator matrix can then be seen as a block matrix.
If we consider the code $\mathcal{C}_{\Fset_2^2}$, then the parameters of the code are $[4,2,3] $.
Notice that   $\mathcal{C}_{\Fset_2}$ is not  MDS, however the code $\mathcal{C}_{\Fset_2^2 }$ is.
\end{example}
 
In general, the minimum distance of $\mathcal{C}_{\mathbb{F}_q}$ is greater than or equal to the minimum distance of  $\mathcal{C}_{\mathbb{F}_{q} ^b}$ (see
\cite{Cardell2012PhDT} for more details).

\subsection{Superregular matrices}
 \begin{definition}
A matrix $A$ is said to be a superregular matrix over $\mathbb{F}_q$ if \textbf{every} square submatrix of $A$ is non-singular over $\mathbb{F}_q$.
\end{definition} 

\begin{example}
    The matrix 
    $$
    A=
    \left[
\begin{array}{ccc}
   1   & 2 & 4 \\
   1   & 3 & 2 \\
   1   & 4 & 2
\end{array}
    \right]    $$
is superregular over 
$\mathbb{F}_7$, since its determinant is nonzero ($\det(A)=2$), all its entries are nonzero, and all its 
$2\times 2$ minors are nonzero over $\mathbb{F}_7$.
\end{example}
  
These matrices are also known as    MDS matrices, due to their relation with MDS (maximum distance separable)  codes as we may see in the following theorem.

\begin{theorem}\cite{Huffman2003bk}
An $[n,k,d]$-linear code $\mathcal{C}$ over $\mathbb{F}_q$ with generator matrix $G=[I_k \mid A]$, where $I_k$ is the $k\times k$  identity matrix and $A\in \mathsf{Mat}_{k \times n-k}(\mathbb{F}_{q})$, is MDS if and only if $A$ is a superregular (MDS) matrix over $\mathbb{F}_q$.
\end{theorem} 

There are several families of matrices that are superregular (as long as the field is large enough), for instance Cauchy and Vandermonde matrices \cite{Roth1985, Burgisser1997bk}. For more details about superregular  matrix constructions, see \cite{Survey} .

Now, we introduce the definition of block superregular matrix.


\begin{definition}
A matrix \( A = [A_{ij}] \in \mathsf{Mat}_{bm \times bt}(\mathbb{F}_{q}) \), where each block \( A_{ij} \) is of size \( b \times b \), is called a  \( b \)-block superregular matrix if  every square submatrix formed by complete \( b \times b \) blocks \( A_{ij} \) is non-singular over \( \mathbb{F}_{q} \).
\end{definition}
 
\begin{example}\label{ex:supblock}
The matrix given by $$  
A=
\left[
\begin{array}{cc|cc}
   1 &  1 &  1 &  1\\
   0 &  1 &  0 &  1\\\hline
   2 &  2 &  3 &  3\\
   0 &  2 &  0 &  3\\
   \end{array}
\right]
$$
is a 2-block superregular matrix over $\mathbb{F}_7$, since the matrices
$$
\left[
\begin{array}{cc}
    1 &  1\\
    0  & 1
\end{array}
\right],
\left[
\begin{array}{cc}
   2 &  2\\
   0  & 2
\end{array}
\right],
 \left[
\begin{array}{cc}
    3 &  3\\
   0  & 3
\end{array}
\right]
$$
and $A$ itself ($\det(A)=1$) are non-singular matrices   over $\mathbb{F}_7$. Notice that $A$ is not superregular over $\mathbb{F}_7$ (it has zero entries).
\end{example}

These matrices are related to MDS $\mathbb{F}_q$-linear codes, as shown in the following theorem.
\begin{theorem}
[\cite{Blaum1999,Cardell2012PhDT}]\label{th:super:MDS}
Let $
  G
  =[     I_{kb} \mid A]  
$
be an $kb \times nb$  generator matrix of an $\mathbb{F}_{q}$-linear code $C_{\mathbb{F}_{q}^{b}}$ with parameters $[n,k]$  over  $\Fset_q^b $.
Then $C_{\mathbb{F}_{q}^{b}}$ is MDS if and only if $A$ is a $b$-block superregular matrix over $\mathbb{F}_q$.
\end{theorem}

In \cite{Bastos2025}, the authors introduced a new construction of superregular block matrices based on the Kronecker product of a superregular matrix and a non-singular matrix. In this work, we consider the 
$\mathbb{F}_q$-linear code generated by one of these matrices, which is presented in the following result.

\begin{theorem}\cite[Corollary 2]{Bastos2025}\label{co:Bi}
    If $A\in \mathsf{Mat}_{ r\times s}(\mathbb{F}_{q})$ is superregular and $ B_1, B_2,..., B_s \in \mathsf{Mat}_{ n\times n}(\mathbb{F}_{q})$ are non-singular, then the matrix  given by
\begin{equation}\label{mat_prod_b2}
M= \left[\begin{array}{cccc}
    a_{11}   B_1 &a_{12}   B_2 &\ldots &a_{1s}   B_s \\
     a_{21}   B_1 &a_{22}   B_2 &\ldots &a_{2s}   B_s \\
     \vdots &\vdots &\ddots&\vdots \\
     a_{r1}   B_1 &a_{r2}   B_2 &\ldots &a_{rs}   B_s \\
\end{array}\right]_{rn \times sn}   
\end{equation}
is $n$-block superregular.
\end{theorem}
Notice that if $B_1 = B_2 = \cdots = B_s = B$, then the matrix $M$ is the Kronecker product of $A$ and $B$, that is, $M = A \otimes B$.

\section{Decoding algorithm in the Erasure Channel}\label{sec:EC}

The $q$-ary erasure channel ($q$-EC) is a discrete memoryless channel with input alphabet $\mathbb{F}_q$ and output alphabet $\mathbb{F}_q\cup\{\varepsilon\}$, where $\varepsilon$ denotes an erasure symbol. Each transmitted symbol   is received correctly with probability $1-\rho$ and is replaced by $\varepsilon$ with probability $\rho$. Therefore, the receiver knows the locations of the erased symbols, although their values remain unknown~\cite{LivaPaoliniChiani2013}.

Let $\mathcal{C}$ be a   $\mathbb{F}_q$-linear code over $\mathbb{F}_q^b$ with  parameters $[n,k]$ and parity check matrix
$$
H=
\left[
M | I_{b(n-k)}
\right],
$$
where
$$
M= A \otimes B=
\left[
\begin{array}{cccc}
a_{11}B_1 & a_{12}B_2 & \ldots &  a_{1k}B_k\\
a_{21}B_1 & a_{22}B_2 & \ldots &  a_{2k}B_k\\
\vdots & \vdots&   &  \vdots\\
a_{n-k\ 1}B_1 & a_{n-k \ 2}B_2 & \ldots &  a_{n-k \ k}B_k\\
\end{array}
\right]_{b(n-k)\times bk},
$$
with $A=[a_{ij}]_{n-k \times k}$
 superregular and $B_1, \ldots, B_k$ non-singular of size $b\times b$. Then, according to Theorem \ref{co:Bi}, $M$
 is $b$-block superregular. 
 This means that $\mathcal{C}$ is an MDS code  over $\mathbb{F}_q^b$ (Theorem~\ref{th:super:MDS}), i.e., the minimum distance of $\mathcal{C}$ is $d=n-k+1$.

 Assume   the codeword $\pmb{c}$ is sent and we receive $\pmb{v}=\pmb{c}+\pmb{e}$, where $\pmb{e}$ is the channel error and assume we are working in the erasure channel, that means that the code can correct up to $n-k$ erasures.

 The syndrome can be computed as the vector
 $$\pmb{s}^t=H\pmb{v}^t=H\pmb{c}^t + H\pmb{e}^t=H\pmb{e}^t.$$
 Consider
 \[ \pmb{s}=[\pmb{s}_1\ \pmb{s}_2 \ \ldots  \ \pmb{s}_{n-k}] \ \ \mbox{and} \ \ \pmb{e}=[\pmb{e}_1\ \pmb{e}_2 \ \ldots \pmb{e}_{n-k} \ \ldots \ \pmb{e}_{n}],  \]
where $\pmb{s}_i, \pmb{e}_j \in \mathbb{F}_q^b$, for $i=1, 2, \ldots, n-k$ and  $j=1, 2,  \ldots, n$.
Then, according to the form of $M$, we have that
 $$
 \pmb{s}_i=\sum_{j=1}^k a_{ij}B_j\pmb{e}_j+\pmb{e}_{i+k}, \quad \text{for } i=1,2, \ldots, n-k.
 $$

Over the erasure channel, the code can decode up to $n-k$ erasures and the receiver knows the position of the erasures. 
Assume we receive the word $\pmb{v}$ and $n-k$ erasures occurred, $t$
 in the information part (the first $k$ components) and $s$ in the parity-check part (the last $n-k$ components), with $t+s=n-k$. 
 Denote by 
 \[  \mathcal{I}=\{i_1, i_2, \ldots, i_t\}\subset \{1,\ldots, k\} \ \ \mbox{and} \ \  \mathcal{J}=\{j_1, j_2, \ldots, j_s\}\subset \{1,\ldots, n-k\},\]
  the position of the erasures in the information and parity-check part, respectively.
 Accordingly, the error vector has the following block structure:
  $$
  \pmb{e}=
  [
  \underbrace{
   0\ 
   \ldots \ 
   \pmb{e}_{i_1}\ 
   \ldots\ 
   \pmb{e}_{i_t}\ 
   \ldots  \  
   0}_{\text{information symbols}} |
   \underbrace{
   0\
   \ldots \
   \pmb{e}_{k+j_1}\
\ldots\
   \pmb{e}_{k+j_s}\
   \ldots\
   0}_{\text{parity-check symbols}}].
  $$
 We need to compute $\pmb{e}_{i_r}$, for $r=1, \ldots, t$, and 
   $\pmb{e}_{k+j_r}$, for $r=1, \ldots, s$.
Notice that:
\begin{align*}
   \pmb{s}^t= H\pmb{e}^t&=
\left[
M | I_{n-k}
\right]\pmb{e}^t
=\left[
\begin{array}{cccc|c}
a_{11}B_1 & a_{12}B_2 & \ldots &  a_{1k}B_k&\multirow{4}{*}{$I_{b(n-k)}$}\\
a_{21}B_1 & a_{22}B_2 & \ldots &  a_{2k}B_k\\
\vdots & \vdots&   &  \vdots\\
a_{n-k, 1}B_1 & a_{n-k,  2}B_2 & \ldots &  a_{n-k, k}B_k\\
\end{array}
\right]\pmb{e}^t \notag\\
&=
\left[
\begin{array}{cc}
    \sum_{r=1}^t a_{1i_r}B_{i_r}\pmb{e}_{i_r}^t \\[5pt] 
    \sum_{r=1}^t a_{2i_r}B_{i_r}\pmb{e}_{i_r}^t\\[5pt] 
        \sum_{r=1}^t a_{3i_r}B_{i_r}\pmb{e}_{i_r}^t\\[5pt] 
    \vdots\\ [5pt]
      \sum_{r=1}^t a_{n-k, i_r}B_{i_r}\pmb{e}_{i_r}^t \label{eq} \\
\end{array}
\right]+
\left[
\begin{array}{c}
   0\\ [-5pt]
   \vdots\\
   \pmb{e}_{k+j_1}^t\\
\vdots\\
   \pmb{e}_{k+j_s}^t\\
   \vdots\\
   0
\end{array}
\right]
\begin{array}{c}
   \phantom{0}\\ 
   \phantom{\vdots}\\
   \leftarrow \text{position } j_1\\
 \vdots \\
  \leftarrow \text{position } j_s\\
   \phantom{\vdots}\\
   \phantom{0}
\end{array}.
\notag\\\end{align*}
This system can also be seen as:

\begin{equation}\label{eq:sistema}
 \pmb{s}_i^t=
 \begin{cases}
     \sum_{r=1}^t a_{i i_r} B_{i_r}\pmb{e}^t_{i_r}+\pmb{e}_{k+i}^t, \quad \text{ if } i\in \mathcal{J}  \\
     \sum_{r=1}^t a_{i i_r} B_{i_r}\pmb{e}^t_{i_r}, \quad \text{ if } i\not \in \mathcal{J} 
 \end{cases},
\end{equation}
for $i=1, 2, \ldots, n-k $.


\color{black}
 Now, denote    $\pmb{f}_r^t=B_{i_r}\pmb{e}_{i_r}^t$,  for $r=1, \ldots, t$, and    $\pmb{f}_{t+r}= \pmb{e}_{k+j_r}$, for $r=1, \ldots s$. 
 Substituting in \eqref{eq:sistema},  we have, in total,  $n-k$ unknowns symbols: $
  \pmb{f}_1, \pmb{f}_2, \ldots, \pmb{f}_{n-k}$.
The system can also be seen as:
\begin{equation}\label{eq:sis:f2}
 \pmb{s}_i^t=
 \begin{cases}
     \sum_{r=1}^t a_{i i_r}  \pmb{f}^t_{r}+\pmb{f}_{t+\ell}^t, \quad \text{ if } i=j_\ell,  \text{ for some  } \ell\ \in \{1,\ldots,s\}  \\
     \sum_{r=1}^t a_{i i_r} \pmb{f}^t_{r}, \quad \text{ if } i \not \in \mathcal{J}
 \end{cases},
\end{equation}
for $i=1, 2, \ldots, n-k, $.
Recall that $\pmb{f}_r\in \mathbb{F}^b_q$, therefore, in total we have $(n-k)b$ unknowns and $(n-k)b$ linear equations in \eqref{eq:sis:f2}. 
Let 
$$
 A^*=
 \left[ 
 \begin{array}{cccc}
   a_{1 i_1}   &  a_{1 i_2}  &\ldots &  a_{1 i_t}   \\
     a_{2 i_1}   &  a_{2 i_2}  &\ldots &  a_{2 i_t}   \\
     \vdots   &   \vdots & &     \vdots   \\
       a_{n-k,i_1}   &  a_{n-k, i_2}  &\ldots &  a_{n-k, i_t}   \\
 \end{array}
 \right]_{n-k\times t},
$$
be  the submatrix of $A$ consisting of the columns indexed by $\mathcal I$, and 
$$
B^*=
[
\pmb{e}_{j_1}\ \pmb{e}_{j_2}\  \ldots \ \pmb{e}_{j_s}
]_{n-k \times s},$$
where $\pmb{e}_j$ denotes the $j$-th standard basis vector of $\mathbb F_2^{n-k}$, i.e., $B^*$ is composed of the columns of the identity matrix  $I_{n-k}$ with indices in $\mathcal{J}$.
Then the matrix of the linear system \eqref{eq:sis:f2} with the unknowns $ \pmb{f}_1, \pmb{f}_2, \ldots, \pmb{f}_{n-k}$ is given by
 $$
  M^*= \left[
\begin{array}{ccccc|c}
   A^*  & O &O&\ldots & O & \multirow{5}{*}{$B^* \otimes I_b$ }\\
  O    &   A^* & O&\ldots & O&  \\
   O & O   &   A^* & \ldots & O & \\
    \vdots & \vdots  &  \vdots &   & \vdots & \\
       O & O   &   O & \ldots & A^*  &\\
\end{array}
   \right]  =
 \left[
 A^* \otimes I_b |  B^* \otimes I_b 
 \right]=[A^* | B^*]\otimes I_b.
 $$

We now analyze whether the matrix $M^*$ is invertible. The matrix $A^*$ 
has size $(n-k)\times t$, while $B^*$ has size $(n-k)\times s$. Since
$t+s=n-k$, the matrix
$
[A^*\mid B^*]
$
is a square matrix of size $(n-k)\times(n-k)$. Consequently,
$
M^*=[A^*\mid B^*]\otimes I_b
$
is also a square matrix. Its determinant is therefore given by
\[
\det(M^*)
=
\det([A^*\mid B^*]\otimes I_b)
=
\det([A^*\mid B^*])^b\det(I_b)^{n-k}
=
\det([A^*\mid B^*])^b.
\]
After a suitable permutation of its rows and columns, the matrix
$[A^*\mid B^*]$ can be written in block triangular form as
\[
\left[
\begin{array}{cc}
A_{\mathcal{J}^c,\mathcal{I}} & 0\\
A_{\mathcal{J},\mathcal{I}} & I_s
\end{array}
\right],
\]
where $A_{\mathcal{J}^c,\mathcal{I}}$ is the submatrix of $A$ obtained
by selecting the rows indexed by $\mathcal{J}^c$ and the columns indexed
by $\mathcal{I}$. Since $|\mathcal{J}^c|=t$, this is a $t\times t$
minor of $A$. As $A$ is superregular, every nontrivial square minor of
$A$ is nonzero. Hence
\[
\det\left(A_{\mathcal{J}^c,\mathcal{I}}\right)\neq 0
\quad \Longrightarrow \quad
\det\left(
\left[
\begin{array}{cc}
A_{\mathcal{J}^c,\mathcal{I}} & 0\\
A_{\mathcal{J},\mathcal{I}} & I_s
\end{array}
\right]
\right)
=
\det(A_{\mathcal{J}^c,\mathcal{I}})\det(I_s)
\neq 0.
\]
Therefore,
$
\det\left([A^*\mid B^*]\right)\neq 0,
$
and consequently $[A^*\mid B^*]$ is non-singular.
Thus, $M^*$ is non-singular, and the linear system \eqref{eq:sis:f2} has a unique solution.
%

\color{black}

In order to obtain the solution, i.e., find the values of, $\pmb{f_1},\pmb{f_2}, \ldots, \pmb{f_{n-k}}$
we have to compute 
$$
\pmb{s}^t=M^*[ \pmb{f_1}\ \pmb{f_2}\  \ldots \ \pmb{f_{n-k}}]^t  
\quad \Longrightarrow\quad 
[ \pmb{f_1}\ \pmb{f_2}\  \ldots \ \pmb{f_{n-k}}]^t=(M^*)^{-1}\pmb{s}^t.
$$
 
Afterwards, we can compute
$
\pmb{e}_{i_r}^t=B_{i_r}^{-1}\pmb{f}_{r}^t$,  for $r=1, \ldots, t$, and   set  $\pmb{e}_{k+j_r}=\pmb{f}_{t+r}$, for $r=1, \ldots s.
$

Algorithm~\ref{alg:erasure} in Appendix A summarizes the decoding process described above, outlining its main steps.

\begin{example}\label{ex:erasure}
Consider the matrices
    $$
    A=
  \left[ 
  \begin{array}{cc}
   1    & 2 \\
   2    & 2
  \end{array}
  \right],
    B_1=
  \left[ 
  \begin{array}{cc}
   1    & 1 \\
   0    & 1
  \end{array}
  \right],
   B_2=
  \left[ 
  \begin{array}{cc}
   1    & 0 \\
   1   & 1
  \end{array}
  \right],
    $$
    where $A$ is superregular and $B_1, B_2$ are non-singular over $\mathbb{F}_3$.
    According to Theorem~\ref{th:super:MDS}, the matrix
    $$
    H=
      \left[ 
  \begin{array}{cc|c}
   B_1    & 2B_2 &\multirow{2}{*}{$I_4$} \\
   2 B_1   & 2B_2&
  \end{array}
  \right]
  =
    \left[ 
  \begin{array}{cccc|c}
   1    & 1 & 2   & 0& \multirow{4}{*}{$I_4$}\\
   0    & 2 &2    & 2&\\
    2   & 2 &2    & 2&\\
     0  & 2 & 0   & 2&\\
  \end{array}
  \right],
    $$
    is the
parity-check matrix of an MDS $\mathbb{F}_3$-linear code with parameters $[4,2,3]$ over $\Fset_3^2$.
    Assume the sent codeword is given by
    $$
    \pmb{c}=[\begin{array}{cc|cc|cc|cc}
       1  &  0 & 1 & 0& 0&1&2&1\\
    \end{array}]
    $$
and that we
    receive the word  
    $$
    \pmb{v}=
    [
    \pmb{v}_1\ \pmb{v}_2 \ \pmb{v}_3\  \pmb{v}_4
    ]
    =
    [\begin{array}{cc|cc|cc|cc}
       \multicolumn{2}{c|}{   \varepsilon \ } & 1 & 0&  \multicolumn{2}{c|}{\  \varepsilon \ }&2&1\\
    \end{array}]. 
    $$
We    assume  that $\pmb{v}_1=\pmb{v}_3=[ 0 \ \ 0]$ to simplify the computations, that is, 
     $$
    \pmb{v}=[\begin{array}{cc|cc|cc|cc}
      0&0 & 1 & 0& 0& 0&2&1\\
    \end{array}].
    $$
    The corresponding syndrome is
    $$
    \left[
    \begin{array}{c}
\pmb{s}_1^t\\[4pt]
\pmb{s}_2^t
    \end{array}
    \right]
    =H\pmb{v}^t=
    \left[ 
    \begin{array}{c}
      2\\
      2\\
      1\\
      0
    \end{array}
    \right]
    $$
    and the error is given by
     $$
    \pmb{e}=[\begin{array}{cc|cc|cc|cc}
      \multicolumn{2}{c|}{ \pmb{e}_1 } & 0 & 0& \multicolumn{2}{c|}{ \pmb{e}_3 } &0&0\\
    \end{array}],
    $$
    with $\pmb{e}_1, \pmb{e}_3\in \mathbb{F}_3^2$.    Now, we know that
    $\pmb{f}_1^t=B_1 \pmb{e}_1^t$,
     $\pmb{f}_2= \pmb{e}_3$, since
     $$ 
     A^*=\left[
     \begin{array}{c}
          1 \\
          2
     \end{array}\right] \quad \mbox{and} \quad 
     B^*=
     \left[
     \begin{array}{c}
          1 \\
          0
     \end{array}\right],
     $$
we have
$$
M^*=
[A^* | B^*] \otimes I_2  =
\left[ 
\begin{array}{cc}
  1 & 1  \\
  2 & 0  
\end{array}\right]
\otimes 
\left[ 
\begin{array}{cc}
  1 & 0   \\
  0 & 1  
\end{array}\right]=
\left[ 
\begin{array}{cc|cc}
  1 & 0 & 1 & 0\\
  0 & 1 &0 & 1\\\hline
  2 & 0& 0& 0\\
0&2&0&0  
\end{array}\right].
$$
Then
$$
    \left[
    \begin{array}{c}
\pmb{s}_1^t\\[4pt]
\pmb{s}_2^t
    \end{array}
    \right]=
\left[
\begin{array}{c}
     2  \\
     2\\
     1\\
     0
\end{array}
\right]
=
\left[ 
\begin{array}{cccc}
  1 & 0 & 1 & 0\\
  0 & 1 &0 & 1\\
  2 & 0& 0& 0\\
0&2&0&0  
\end{array}\right]
\left[
\begin{array}{c}
\pmb{f}_1^t\\[4pt]
\pmb{f}_2^t
\end{array}
\right] 
$$
and
$$
    \left[
    \begin{array}{c}
\pmb{f}_1^t\\[4pt]
\pmb{f}_2^t
    \end{array}
    \right]
=
\left[ 
\begin{array}{cccc}
  1 & 0 & 1 & 0\\
  0 & 1 &0 & 1\\
  2 & 0& 0& 0\\
0&2&0&0  
\end{array}\right]^{-1}
\left[
\begin{array}{c}
     2  \\
     2\\
     1\\
     0
\end{array}
\right]=
\left[
\begin{array}{cccc}
  0 & 0 & 2 & 0\\
  0 & 0 &0 & 2\\
  1 & 0& 1& 0\\
0&1&0&1  
\end{array}\right] 
\left[
\begin{array}{c}
     2  \\
     2\\
     1\\
     0
\end{array}
\right]
=
\left[ 
\begin{array}{c}
2\\
0\\
0\\
2
\end{array}\right].
$$
Therefore $\pmb{f}_1=[2 \ 0 ]$, $\pmb{f}_2=[0\ 2]$ and, as a consequence,
$$
\pmb{e}_1^t=B_1^{-1}\pmb{f}_1^t=
\left[
\begin{array}{cc}
    1 &2  \\
     0&1 
\end{array}
\right]
\left[
\begin{array}{c}
     2  \\
     0 
\end{array}
\right]
=
\left[
\begin{array}{c}
     2  \\
     0 
\end{array}
\right]
\quad \text{and}  \quad \pmb{e}_3=\pmb{f}_2=
\left[
\begin{array}{c}
     0  \\
     2 
\end{array}
\right].
$$
The sent codeword is:
\begin{align*}
    \pmb{c}=& \pmb{v}-\pmb{e}\\
           =& [\begin{array}{cc|cc|cc|cc}
      0&0 & 1 & 0& 0& 0&2&1\\
    \end{array}] 
    -  \begin{array}{cc|cc|cc|cc}
      [\pmb{2}&\pmb{0} & 0 & 0& \pmb{0}& \pmb{2}&0&0\\
          \end{array}]\\
      = &[\begin{array}{cc|cc|cc|cc}
      1&0 & 1 & 0& 0& 1&2&1\\
    \end{array}].
\end{align*}
\end{example}

 Notice that in Algorithm~\ref{alg:erasure} we do not  need all matrices $B_i^{-1}$, 
just the ones corresponding to the positions of the errors in $\mathcal{I}$.
 Furthermore, when $M=A\otimes B$, we have a special case when $B=B_1=\cdots=B_k$. In this case, the algorithm is less complex, since we only have to compute $B^{-1}$ once.

If we focus on the inverse of $M^*$, we see that 
$$
(M^*)^{-1}=[A^*\mid B^*]^{-1}\otimes I_b.
$$
Therefore, if the inverse of $[A^*\mid B^*]$ can be efficiently computed, there is no need to explicitly compute the inverse of the larger matrix $M^*$. In this way, the decoding algorithm avoids the additional computations  inverting a matrix of size $(n-k)b\times(n-k)b$.

Let $\mathcal{J}^c=\{r_1,\ldots,r_t\}$ denote the complement of $\mathcal{J}$ in $\{1,\ldots,n-k\}$. Since $B^*$ consists of the columns of $I_{n-k}$ indexed by $\mathcal{J}$, after a suitable permutation of the rows and columns, the matrix $[A^*\mid B^*]$ can be written as
$$
\begin{bmatrix}
A_{\mathcal{J}^c,\mathcal{I}} & 0\\
A_{\mathcal{J},\mathcal{I}} & I_s
\end{bmatrix}.
$$
Let $C=A_{\mathcal{J}^c,\mathcal{I}}$. Then
$$
\begin{bmatrix}
C & 0\\
A_{\mathcal{J},\mathcal{I}} & I_s
\end{bmatrix}^{-1}
=
\begin{bmatrix}
C^{-1} & 0\\
-A_{\mathcal{J},\mathcal{I}}C^{-1} & I_s
\end{bmatrix}.
$$
Therefore, up to the corresponding permutation matrices, the inverse of $[A^*\mid B^*]$ is determined by the inverse of the $t\times t$ submatrix $A_{\mathcal{J}^c,\mathcal{I}}$.

\begin{example}
    Consider de MDS array code presented in  Example~\ref{ex:erasure}, where,
$$
[A^*\mid B^*]
=
\left[
\begin{array}{cc}
1&1\\
2&0
\end{array}
\right].
$$
Its inverse over $\mathbb{F}_3$ is
$$
[A^*\mid B^*]^{-1}
=
\left[
\begin{array}{cc}
0&2\\
1&1
\end{array}
\right].
$$
Therefore,
$$
(M^*)^{-1}
=
[A^*\mid B^*]^{-1}\otimes I_2
=
\left[
\begin{array}{cc}
0&2\\
1&1
\end{array}
\right]
\otimes
\left[
\begin{array}{cc}
1&0\\
0&1
\end{array}
\right]=
(M^*)^{-1}
=
\left[
\begin{array}{cc|cc}
0&0&2&0\\
0&0&0&2\\ \hline
1&0&1&0\\
0&1&0&1
\end{array}
\right].
$$
This example is particularly simple since $[A^*\mid B^*]$ is already in triangular form, convenient to compute the inverse. For larger matrices, the columns corresponding to $B^*$ may be distributed among different positions. In this case, suitable row and column permutations can be applied to obtain a block triangular form, after which the inverse can be computed using the inverse of the corresponding submatrix of $A$.
\end{example}

\color{black}

\section{Decoding algorithm no Symmetric Channel}

 The $q$-ary symmetric channel ($q$-SC) is a discrete memory-less channel that generalizes the binary symmetric channel to the alphabet $\mathbb{F}_q$ \cite{Weidmann2012}.
 Its input and output alphabets are
both identified with $\mathbb{F}_q$, and its transition probabilities
are given by
\[
\Pr(Y=y\mid X=x)
=
\begin{cases}
1-p, & y=x\\[2mm]
\dfrac{p}{q-1}, & y\neq x
\end{cases}.
\]
Thus, each transmitted symbol is received correctly with probability
$1-p$, whereas, with probability $p$, it is replaced by one of the
remaining $q-1$ symbols, each occurring with equal probability
$p/(q-1)$~\cite{WeidmannLechner2012}. For $q=2$, this
model reduces to the binary symmetric channel.

If a $q$-ary code has minimum Hamming distance $d$, then
nearest-neighbor decoding with respect to the Hamming distance corrects
every error pattern of Hamming weight at most
\[
t
=
\left\lfloor\frac{d-1}{2}\right\rfloor.
\]
Since $\mathcal{C}$ is an $[n,k,n-k+1]_q$ MDS code, it can correct up to
\[
t
=
\left\lfloor\frac{n-k}{2}\right\rfloor
\]
symbol errors over the $q$-ary symmetric channel.

Consider again the $\mathbb{F}_q$-linear code 
$\mathcal{C}$, with parameters $[n,k,n-k+1]$  consider in Section~\ref{sec:EC}.
 Assume the codeword  $ \pmb{c}$ is sent and we receive $\pmb{v}=\pmb{c}+\pmb{e}$.
 The syndrome can be computed as the vector
 $$\pmb{s}^t=H\pmb{v}^t=H\pmb{c}^t + H\pmb{e}^t=H\pmb{e}^t.$$
 Consider
 \[ \pmb{s}=[\pmb{s}_1\ \pmb{s}_2 \ \ldots  \ \pmb{s}_{n-k}] \quad \mbox{and} \quad
        \pmb{e}=[\pmb{e}_1\ \pmb{e}_2 \ \ldots \pmb{e}_{n-k} \ \ldots \ \pmb{e}_{n}], \]
where $\pmb{s}_i, \pmb{e}_j \in \mathbb{F}_q^b$, for $i=1, 2, \ldots, n-k$ and  $j=1, 2,  \ldots, n$.

The decoding procedure is developed by considering the possible locations of the symbol errors, leading to the cases presented in the following subsections.

\subsection{Errors in the parity symbols}

If $l$ symbol errors $\pmb{e}_{j_1},\ldots,\pmb{e}_{j_l}\neq\pmb{0}$ occur in the parity symbols with indices $k+j_1,\ldots,k+j_l$, respectively, where $1\leq j_1<\cdots<j_l\leq n-k$, then the error vector is given by
 $$
  \pmb{e}=[\underbrace{\pmb{0}\  \ldots\pmb{0}  }_{k \text{ symbols}}\underbrace{\pmb{0}\ \ldots\ \pmb{e}_{j_1}\ \ldots\ 
\pmb{e}_{j_2}\ \ldots\ 
\pmb{e}_{j_l}\ \ldots\ \pmb{0}}_{n-k \text{ symbols }} ]\\
 $$ 
and the resultant syndrome gives all $l$ errors in their corresponding positions:

\begin{align*}
\pmb{s}^t
&=
H\pmb{e}^t
=
\left[
M \mid I_{b(n-k)}
\right]\pmb{e}^t
\\
&=
\left[
\begin{small}
\begin{array}{cccc|c}
a_{11}B_1 & a_{12}B_2 & \ldots & a_{1k}B_k & \multirow{4}{*}{$I_{b(n-k)}$} \\
a_{21}B_1 & a_{22}B_2 & \ldots & a_{2k}B_k & \\
\vdots & \vdots & & \vdots & \\
a_{n-k,1}B_1 & a_{n-k,2}B_2 & \ldots & a_{n-k,k}B_k &
\end{array}
\end{small}
\right]
\left[
\begin{small}
\begin{array}{c}
\pmb{0} \\
\vdots \\
\pmb{0} \\
\pmb{e}_{j_1}^t \\
\vdots \\
\pmb{e}_{j_2}^t \\
\vdots \\
\pmb{e}_{j_l}^t \\
\vdots \\
\pmb{0}
\end{array}
\end{small}
\right]
\\
&=
\left[
\begin{array}{c}
\pmb{s}_1^t \\
\vdots \\
\pmb{s}_{j_1}^t \\
\vdots \\
\pmb{s}_{j_2}^t \\
\vdots \\
\pmb{s}_{j_l}^t \\
\vdots \\
\pmb{s}_{n-k}^t
\end{array}
\right]
=
\left[
\begin{array}{c}
\pmb{0} \\
\vdots \\
\pmb{e}_{j_1}^t \\
\vdots \\
\pmb{e}_{j_2}^t \\
\vdots \\
\pmb{e}_{j_l}^t \\
\vdots \\
\pmb{0}
\end{array}
\right].
\end{align*}
Thus, for every $j=1,\ldots,n-k$,
\[
\pmb{s}_j=
\begin{cases}
\pmb{e}_j, & j\in\{j_1,\ldots,j_l\}\\
\pmb{0}, & j\notin\{j_1,\ldots,j_l\}
\end{cases}.
\]
Notice that $\pmb{e}_{j_i}$ appears in the $k+j_i$-th position in the error vector (whose length is $n$) and in the $j_i$-th position in the syndrome vector (whose length is $n-k$), for $i=1,\ldots,l$. Hence, the syndrome directly reveals the locations and values of all $l$ symbol errors occurring in the parity symbols.

 \begin{example}\label{exeerroparidade}
 Consider the matrices
    $$
    A=
  \left[ 
  \begin{array}{cc}
   1    & 2 \\
   2    & 2
  \end{array}
  \right],
    B_1=
  \left[ 
  \begin{array}{cc}
   1    & 1 \\
   0    & 1
  \end{array}
  \right],
   B_2=
  \left[ 
  \begin{array}{cc}
   1    & 0 \\
   1   & 1
  \end{array}
  \right],
    $$
    where $A$ is superregular and $B_1, B_2$ are non-singular over $\mathbb{F}_3$.
    The corresponding matrix
    $$
    H=
      \left[ 
  \begin{array}{cc|c}
   B_1    & 2B_2 &\multirow{2}{*}{$I_4$} \\
   2 B_1   & 2B_2&
  \end{array}
  \right]
  =
    \left[ 
  \begin{array}{cccc|c}
   1    & 1 & 2   & 0& \multirow{4}{*}{$I_4$}\\
   0    & 1 &2    & 2&\\
    2   & 2 &2    & 2&\\
     0  & 2 & 0   & 2&\\
  \end{array}
  \right],
    $$
is the parity-check matrix of an MDS $\mathbb{F}_3$-linear code with parameters $[4,2,3]$ over $\mathbb{F}_3^2$.
This code has 2 parity symbols and 2 information symbols.  
    Assume the sent codeword is given by
    $$
    \pmb{c}=[\begin{array}{cc|cc|cc|cc}
       1  &  0 & 1 & 0& 0&1&2&0\\
    \end{array}]
    $$
and that we
    receive the word
    $$
    \pmb{v}=[\begin{array}{cc|cc|cc|cc}
       1&0 & 1 & 0&   \pmb{0} & \pmb{0}&2&0\\
    \end{array}]. 
    $$
    The syndrome is calculated by
    $$
    \pmb{s}^t=H\pmb{v}^t=
    \left[ 
    \begin{array}{c}
      0\\
      2\\
      0\\
      0
    \end{array}
    \right].
    $$
Thus, 
    $$
    \pmb{s}_1^t=\left[
    \begin{array}{c}
    0 \\
    2
    \end{array}
    \right]
    \quad \mbox{and} \quad
    \pmb{s}_2^t=\left[
    \begin{array}{c}
    0 \\
    0
    \end{array}
    \right].
    $$
    Since the error appears only in the 1st position of the vector syndrome, then 
     the error is in the first parity symbol, that is, in the $3$rd  position of the codeword (2 information symbols + 1 parity symbol). 
     Moreover, the error is equal to the syndrome symbol $\pmb{s}_1$.
    Therefore, the codeword is:
  \begin{align*}
   \pmb{c}=
  &[\begin{array}{cc|cc|cc|cc}
1 &0& 1& 0& 0& 0& 2& 0\\
  \end{array}]-
  [\begin{array}{cc|cc|cc|cc}
0 &0& 0& 0& \pmb{0}& \pmb{2}& 0& 0\\
  \end{array}]\\
  =
  &[\begin{array}{cc|cc|cc|cc}
1 &0& 1& 0& 0& 1& 2& 0\\
  \end{array}].
  \end{align*}
 \end{example}

 \subsection{One error in the information symbols}
 If one symbol error $\pmb{e}_j\not = \pmb{0}$ occurs in the information part,  with $1\le j \le k$), the   error vector is given by
 $$
  \pmb{e}=[\underbrace{\pmb{0}\ \pmb{0}  \ \ldots\pmb{0} \ \pmb{e}_{j} \ \pmb{0} \ldots \ \pmb{0}}_{k \text{ symbols}}\ | \underbrace{\pmb{0} \ldots\pmb{0}}_{
  \begin{array}{c}
  n-k\\ 
  \text{ \footnotesize symbols }\\\end{array}} ]\\
 $$
 and the resultant syndrome is given by:
\begin{align*}
 \pmb{s}^t&=H\pmb{e}^t =
\left[
M | I_{b(n-k)}
\right]\pmb{e}^t
\\
&=\left[
\begin{array}{cccc|c}
a_{11}B_1 & a_{12}B_2 & \ldots &  a_{1k}B_k&\multirow{4}{*}{$I_{b(n-k)}$}\\
a_{21}B_1 & a_{22}B_2 & \ldots &  a_{2k}B_k&\\
\vdots & \vdots&   &  \vdots&\\
a_{n-k, 1}B_1 & a_{n-k, 2}B_2 & \ldots &  a_{n-k, k}B_k&\\
\end{array}
\right]
\left[
\begin{array}{c}
\pmb{0}\\
\vdots\\
\pmb{e}_j^t\\
\vdots\\
\pmb{0}\\
\end{array}
\right]
=
\left[
\begin{array}{c}
a_{1j}B_j\pmb{e}_j^t\\
a_{2j}B_j\pmb{e}_j^t\\
\vdots\\
a_{n-k, j}B_j\pmb{e}_j^t\\
\end{array}
\right].
\end{align*}
 If we multiply each symbol of the syndrome by the inverse of each $B_j$, that is, we multiply $I_{n-k}\otimes B_j^{-1}$ by  $\pmb{s}^t$ we obtain
 \begin{align}\label{eq:1}
 (I_{n-k}\otimes B_j^{-1})\pmb{s}^t =
\left[ 
\begin{array}{ccccc}
  B_j^{-1}   &  & & &  \\
             &   B_j^{-1}  & & &  \\
      &  &\ddots & &  \\
       &  & &  & B_j^{-1}    \\
\end{array}
\right]
\left[
\begin{array}{c}
a_{1j}B_j\pmb{e}_j^t\\
a_{2j}B_j\pmb{e}_j^t\\
\vdots\\
a_{n-k, j}B_j\pmb{e}_j^t\\
\end{array}
\right]
&
&=
\left[
\begin{array}{c}
a_{1j}\pmb{e}_j^t\\
a_{2j}\pmb{e}_j^t\\
\vdots\\
a_{n-k, j}\pmb{e}_j^t\\
\end{array}
\right] 
=A_j\otimes  \pmb{e}_j^t,
  \end{align}
where $A_j$ is the    the $j$-th column of $A$.

 The complete decoding procedure is presented in
Algorithm~\ref{alg:one-info-distinct-B} in Appendix~A.

 \begin{example}
 Consider the same code as in Example \ref{exeerroparidade}.
    Assume the sent codeword is given by
    $$
    \pmb{c}=[\begin{array}{cc|cc|cc|cc}
       1  &  0 & 1 & 0& 0&1&2&0\\
    \end{array}]
    $$
and that we
    receive the word
    $$
    \pmb{v}=[\begin{array}{cc|cc|cc|cc}
       \pmb{0}&\pmb{0} & 1 & 0&   0 & 1&2&0
    \end{array}]. 
    $$
  The syndrome is calculated by
    $$
    \pmb{s}^t=H\pmb{v}^t=
    \left[ 
    \begin{array}{c}
      2\\
      0\\
      1\\
      0
    \end{array}
    \right],
    $$
that is, 
    $$
    \pmb{s}_1=\left[
        2 \
    0
        \right]
    \quad \text{and} \quad
    \pmb{s}_2=\left[
    1 \
    0
    \right].
    $$
Since all the syndrome symbols are affected, this means that the error is in one of the $2$ information symbols. 
       Now, we consider the inverses of $B_{1} $ and $B_{2} $:

       \[
B_{1}^{-1}=
\left[
\begin{array}{cc}
    1 & 2 \\
    0 & 1
\end{array}
\right]   
\quad
\text{and}
\quad 
B_{2}^{-1}=
\left[
\begin{array}{cc}
    1 & 0 \\
    2 & 1
\end{array}
\right].    
       \]
  Now we multiply each symbol of the syndrome by both inverses:
  
\begin{minipage}[t]{0.48\textwidth}
\[
\begin{aligned}
B_1^{-1}\pmb{s}_1^t
&=
\left[
\begin{array}{cc}
1 & 2\\
0 & 1
\end{array}
\right]
\left[
\begin{array}{c}
2\\
0
\end{array}
\right]
=
\left[
\begin{array}{c}
2\\
0
\end{array}
\right],
\\[2mm]
B_1^{-1}\pmb{s}_2^t
&=
\left[
\begin{array}{cc}
1 & 2\\
0 & 1
\end{array}
\right]
\left[
\begin{array}{c}
1\\
0
\end{array}
\right]
=
\left[
\begin{array}{c}
1\\
0
\end{array}
\right],
\end{aligned}
\]
\end{minipage}
\hfill
\begin{minipage}[t]{0.48\textwidth}
\[
\begin{aligned}
B_2^{-1}\pmb{s}_1^t
&=
\left[
\begin{array}{cc}
1 & 0\\
2 & 1
\end{array}
\right]
\left[
\begin{array}{c}
2\\
0
\end{array}
\right]
=
\left[
\begin{array}{c}
2\\
1
\end{array}
\right],
\\[2mm]
B_2^{-1}\pmb{s}_2^t
&=
\left[
\begin{array}{cc}
1 & 0\\
2 & 1
\end{array}
\right]
\left[
\begin{array}{c}
1\\
0
\end{array}
\right]
=
\left[
\begin{array}{c}
1\\
2
\end{array}
\right].
\end{aligned}
\]
\end{minipage}
 \medskip
According to \eqref{eq:1}, for some $j\in\{1,\ldots,k\}$, we have
\[
(I_2\otimes B_j^{-1})\pmb{s}^t=A_j\otimes \pmb{e}_j^t.
\]
Therefore, we only need to determine the corresponding index $j$.
For $j=1$, we have
       

       \[
B_1^{-1}  \pmb{s}_1^t 
=
 \left[
\begin{array}{c}
2\\
0
\end{array}
\right]   
=
\textcolor{red}{1}
\left[
\begin{array}{c}
2\\
0
\end{array}
\right] 
=
\textcolor{blue}{2}
\left[
\begin{array}{c}
1\\
0
\end{array}
\right]
      \quad \mbox{and} \quad
B_1^{-1}  \pmb{s}_2^t=
 \left[
\begin{array}{c}
1\\
0
\end{array}
\right] =
\textcolor{red}{2}
\left[
\begin{array}{c}
2\\
0
\end{array}
\right]
=
\textcolor{blue}{1}
\left[
\begin{array}{c}
1\\
0
\end{array}
\right] .
       \]
     Since 
\textcolor{blue}{$\begin{bmatrix}
     2  \\
     1 
\end{bmatrix}$}
 is not a column of 
$A$ but 
\textcolor{red}{$\begin{bmatrix}
     1  \\
     2 
\end{bmatrix}$}
 is the first column of $A$, the only possibility is that the error occurred in the first information symbol and the error   is $[2\ 0 ]$.
 In this case the codeword would be:
   \begin{align*}
  \pmb{c}_1=
  &[\begin{array}{cc|cc|cc|cc}
0 &0& 1& 0& 0& 1& 2& 0\\
  \end{array}]- [\begin{array}{cc|cc|cc|cc}
\mathbf{2} &\mathbf{0}& 0& 0& 0& 0& 0& 0\\
  \end{array}]\\
  =
  &[\begin{array}{cc|cc|cc|cc}
1 &0& 1& 0& 0& 1& 2& 0\\
  \end{array}].
  \end{align*}
Now, for $j=2$, we have
       \[
B_2^{-1}  \pmb{s}_1^t 
=
 \left[
\begin{array}{c}
2\\
1
\end{array}
\right]   
=
\textcolor{blue}{1}
\left[
\begin{array}{c}
2\\
1
\end{array}
\right] 
=
\textcolor{red}{2}
\left[
\begin{array}{c}
1\\
2
\end{array}
\right] 
     \quad \mbox{and} \quad
B_2^{-1}  \pmb{s}_2^t=
 \left[
\begin{array}{c}
2\\
1
\end{array}
\right] =
\textcolor{blue}{1}
\left[
\begin{array}{c}
2\\
1
\end{array}
\right]
=
\textcolor{red}{2}
\left[
\begin{array}{c}
1\\
2
\end{array}
\right] .\]
    Since 
\textcolor{blue}{$\begin{bmatrix}
     1  \\
     1 
\end{bmatrix}$}
 is not a column of 
$A$ but 
\textcolor{red}{$\begin{bmatrix}
     2  \\
     2 
\end{bmatrix}$}
 is the second column of $A$, the only possibility is that the error occurred in the second information symbol and the error  is $[1\ 2 ]$.
 In this case the codeword would be:
   \begin{align*}
   \pmb{c}_2=
  &[\begin{array}{cc|cc|cc|cc}
0 &0& 1& 0& 0& 1& 2& 0\\
  \end{array}]-   [\begin{array}{cc|cc|cc|cc}
0 &0& \mathbf{1}& \mathbf{2}& 0& 0& 0& 0\\
  \end{array}]\\
  =
  &[\begin{array}{cc|cc|cc|cc}
0 &0& 0& 1& 0& 1& 2& 0\\
  \end{array}].
  \end{align*}    
         To check which word is correct, we multiply again by $H$ in order to obtain the syndrome.
         It is possible to check that
         $$
         H \pmb{c}_1^t=
         \left[
\begin{array}{c}
   0\\
   0\\
   0\\
   0
\end{array}
         \right]
         \quad \text{and} \quad
                  H \pmb{c}_2^t=
         \left[
\begin{array}{c}
  0\\
   0\\
   1\\
   2\\
\end{array}
         \right].
         $$       
       Therefore, the error occurs in the $1$st information symbol and it is given by $\pmb{e_1}=[2\ 0]$. 
The codeword is
  \begin{align*}
   \pmb{c}_1=
  &[\begin{array}{cc|cc|cc|cc}
1 &0& 1& 0& 0& 1& 2& 0\\
  \end{array}].
  \end{align*}


 \end{example}

   \begin{remark}
If we consider $B_1=B_2=\cdots = B_k=B$, that is all, the $B_i$ matrices are equal, the complexity of the algorithm is lower, since we only have to compute one inverse matrix, $B^{-1}$ and we do not have to test several possible cases. 
Algorithm~\ref{alg:one-info-equal-B} presents the recovery procedure for the special case in which
all the matrices $B_j$ are equal, that is, $B_1=\cdots=B_k=B$.
 \end{remark}
 \begin{example}
 
Consider the matrices
    $$
    A=
  \left[ 
  \begin{array}{cc}
   1    & 2 \\
   2    & 2
  \end{array}
  \right] \quad \text{and}
 \quad     B=
  \left[ 
  \begin{array}{cc}
   1    & 1 \\
   0    & 1
  \end{array}
  \right],
    $$
    where $A$ is superregular and $B$ is non-singular over $\mathbb{F}_3$.
    The corresponding matrix
    $$
    H=
      \left[ 
  \begin{array}{cc|c}
   B    & 2B &\multirow{2}{*}{$I_4$} \\
   2 B   & 2B&
  \end{array}
  \right]
  =
    \left[ 
  \begin{array}{cccc|c}
   1    & 1 & 2   & 2& \multirow{4}{*}{$I_4$}\\
   0    & 1 &0    & 2&\\
    2   & 2 &2    & 2&\\
     0  & 2 & 0   & 2&\\
  \end{array}
  \right],
    $$
is the parity-check matrix of an MDS $\mathbb{F}_3$-linear code with parameters $[4,2,3]$ over $\mathbb{F}_3^2$.
     Assume the sent codeword is given by
    $$
    \pmb{c}=[\begin{array}{cc|cc|cc|cc}
        1  & 0 &  1 &  0 &  0 &  0 &  2 &  0
    \end{array}]
    $$
and that we
    receive the word
    $$
    \pmb{v}=[\begin{array}{cc|cc|cc|cc}
       1&0 & \pmb{0} & \pmb{1}&   0 & 0&2&0\\
    \end{array}]. 
    $$
   The syndrome is calculated by
    $$
    \pmb{s}^t=H\pmb{v}^t=
    \left[ 
    \begin{array}{c}
      0\\
      2\\
      0\\
      2
    \end{array}
    \right].
    $$
    Since all the syndrome symbols are affected, we know the error occurs in one of the first $2$ information symbols.
 Now, we multiply each symbol of the syndrome by $B^{-1}$:
    \[
B^{-1} \pmb{s}_1=
\left[
\begin{array}{cc}
    1 & 2 \\
    0 & 1
\end{array}
\right]   
\left[
\begin{array}{c}
0\\
2
\end{array}
\right]  =
 \left[
\begin{array}{c}
1\\
2
\end{array}
\right] =
2
\left[
\begin{array}{c}
2\\
1
\end{array}
\right] 
       \quad \mbox{and} \quad
B^{-1} \pmb{s}_2=
\left[
\begin{array}{cc}
    1 & 2 \\
    0 & 1
\end{array}
\right]   
\left[
\begin{array}{c}
0\\
2
\end{array}
\right]  =
 \left[
\begin{array}{c}
1\\
2
\end{array}
\right] =
2
\left[
\begin{array}{c}
2\\
1
\end{array}
\right]. 
     \]
       When we multiply the syndrome by the inverse of $B$ we obtain $A_j \otimes e_j$.
       In this case $j=2$ and the error is $[2\ 1]$.
       Therefore the error is in the second information symbol and the sent codeword is:
         \begin{align*}
   \pmb{c}=
  &[\begin{array}{cc|cc|cc|cc}
1 &0& 0& 1& 0& 0& 2& 0\\
  \end{array}]-[\begin{array}{cc|cc|cc|cc}
0 &0& \pmb{2}& \pmb{1}& 0& 0& 0& 0\\
  \end{array}]\\
  =
  &[\begin{array}{cc|cc|cc|cc}
1 &0& 1& 0& 0& 0& 2& 0\\
  \end{array}].
  \end{align*}
       
    \end{example}

 \begin{remark}
To improve the efficiency of the algorithm, we can choose $B$ matrices whose inverse is easy to compute, such as diagonal,   orthogonal, Hadamard-type, permutation, triangular, Householder, circulant, or block-diagonal matrices. 
\end{remark}

 \subsection{One error in the information symbols and one error in the parity symbols}
 Consider two symbol errors, 
 one error, $\pmb{e}_{j_1}\not = \pmb{0}$,   in the information part and the other, $e_{j_2}\not = \pmb{0}$, in the parity part, with $1\le j_1\leq k$ and  $1\leq j_2 \leq n-k$.
The error vector  is given by
 $$
  \pmb{e}=[\underbrace{\pmb{0}\ \pmb{0}  \ \ldots\pmb{0} \ \pmb{e}_{j_1} \ \pmb{0} \ldots \ \pmb{0}}_{k \text{ symbols}}\ | \ \underbrace{\pmb{0} \ldots\pmb{0}\ \pmb{e}_{j_2}\ \pmb{0}\ldots \pmb{0}}_{n-k \text{ symbols }} ]\\
 $$
 and the resultant syndrome is given by:

\begin{align*}
 \pmb{s}^t=H\pmb{e}^t&=
\left[
M | I_{b(n-k)}
\right]\pmb{e}^t
\\
&=\left[
\begin{array}{cccc|c}
a_{11}B_1 & a_{12}B_2 & \ldots &  a_{1k}B_k&\multirow{4}{*}{$I_{b(n-k)}$ }\\
a_{21}B_1 & a_{22}B_2 & \ldots &  a_{2k}B_k&\\
\vdots & \vdots&   &  \vdots&\\
a_{n-k, 1}B_1 & a_{n-k, 2}B_2 & \ldots &  a_{n-k, k}B_k&\\
\end{array}
\right]
\left[
\begin{array}{c}
\pmb{0}\\
\vdots\\
\pmb{e}_{j_1}^t\\
\vdots\\
\pmb{0}\\
\pmb{e}_{j_2}^t\\
\pmb{0}\\
\vdots\\
\pmb{0}\\
\end{array}
\right]\\
&=
\left[
\begin{array}{c}
a_{1j_1}B_{j_1}\pmb{e}_{j_1}^t \\
a_{2j_1}B_{j_1}\pmb{e}_{j_1}^t \\
\vdots\\
a_{j_2 j_1}B_{j_1}\pmb{e}_{j_1}^t+\pmb{e}_{j_2}^t\\
\vdots\\
a_{n-k, j_1}B_{j_1}\pmb{e}_{j_1}^t \\
\end{array}
\right].
\end{align*}

 If we multiply each symbol of the syndrome on the left by the inverse of  $B_{j_1}$, that is, we multiply  $I_{n-k}\otimes B_{j_1}^{-1}$ by  $\pmb{s}^t$, we obtain
\begin{equation}\label{eq:doserros}
 (I_{n-k}\otimes B_{j_1}^{-1})\pmb{s}^t =
\left[ 
\begin{array}{ccccc}
  B_{j_1}^{-1}   &  & & &  \\
             &   B_{j_1}^{-1}  & & &  \\
      &  &\ddots & &  \\
       &  & &  & B_{j_1}^{-1}    \\
\end{array}
\right]
\left[
\begin{array}{c}
a_{1j_1}B_{j_1}\pmb{e}_{j_1}^t \\
a_{2j_1}B_{j_1}\pmb{e}_{j_1}^t \\
\vdots\\
a_{j_2 j_1}B_{j_1}\pmb{e}_{j_1}^t+\pmb{e}_{j_2}^t\\
\vdots\\
a_{n-k, j_1}B_{j_1}\pmb{e}_{j_1}^t \\
\end{array}
\right]
=
\left[
\begin{array}{c}
a_{1j_1}\pmb{e}_{j_1}^t \\
a_{2j_1} \pmb{e}_{j_1}^t \\
\vdots\\
a_{j_2 j_1} \pmb{e}_{j_1}^t+B_{j_1}^{-1}\pmb{e}_{j_2}^t\\
\vdots\\
a_{n-k, j_1} \pmb{e}_{j_1}^t \\
\end{array}
\right].
 \end{equation}

Now, since the matrix $A$ is superregular, all entries are different from zero and we can define $D=diag(a_{1{j_1}}^{-1}, a_{1{j_2}}^{-1}, \ldots, a_{n-k,{j_1}}^{-1}) $.
Thus, we can multipy
 \eqref{eq:doserros} 
 by $D  \otimes I_b$:
\begin{align*}
   (D \otimes I_b)
    (I_{n-k}\otimes B_{j_1}^{-1})\pmb{s}^t=(
    D  \otimes B_{j_1}^{-1} ) \pmb{s}^t
=
\left[
\begin{array}{c}
 \pmb{e}_{j_1}^t \\
  \pmb{e}_{j_1}^t \\
\vdots\\
 \pmb{e}_{j_1}^t+a_{j_2 j_1}^{-1}B_{j_1}^{-1}\pmb{e}_{j_2}^t\\
\vdots\\
  \pmb{e}_{j_1}^t \\
\end{array}
\right].
\end{align*}

 Notice that, in the resultant vector,    all the syndrome symbols  are equal to $\pmb{e}_{j_1}^t$ except for the $j_2$-th symbol, which corresponds to the error $\pmb{e}_{j_2}$.
Therefore, we can isolate $\pmb{e}_{j_2}$ and obtain it directly.

 The complete decoding procedure is presented in
Algorithm~\ref{alg:one-information-one-parity-error} in Appendix~A.
 \begin{example}\label{exemploF13}
Consider the superregular matrix in $\mathbb{F}_{13}$ given by:
$$
A=
\left[
\begin{array}{ccc}
   6 &  9  &    6\\
   4  & 3 &    1\\
   3  & 3&      9\\
\end{array}
\right]
$$
and the non-singular matrix
$$
B=
\left[
\begin{array}{cc}
    1 & 1 \\
  0   & 1
\end{array}
\right].
$$
The corresponding matrix
$$
H=
[A\otimes B\ | \ I_6]=
\left[
\begin{array}{ccc|c}
   6B &  9B  &    6B&\multirow{3}{*}{$I_6$}\\
   4B  & 3B &    1B\\
   3B  & 3B&      9B\\
\end{array}
\right],
$$
is the parity check matrix
of an $[6,3,4]$ MDS $\mathbb{F}_{13}$-linear code over $\mathbb{F}_{13}^2$.
Assume that the  codeword
$$
 \pmb{c}=[
\begin{array}{cc|cc|cc|cc|cc|cc}
1 &   0&    1&    0&    0&    0&   11&    0&    6&    0&    7&     0
\end{array}]
$$
is sent, but we receive
$$ \pmb{v}=[
\begin{array}{cc|cc|cc|cc|cc|cc}
1 &   0&   \pmb{1}&    \pmb{1}&    0&    0&   11&    0&    6&    0&    \pmb{7}&    \pmb{1}].
\end{array}$$
The syndrome is calculated by
$$
\pmb{s}^t=
H \pmb{v}^t=
\left[
\begin{array}{c}
 9\\
   9\\
   3\\
   3\\
   3\\
   4
   \end{array}
   \right],
   $$
   that is,
   $$
   \pmb{s}_1 =[9\ 9],\ \ 
    \pmb{s}_2 =[ 3\ 3] \quad \text{and}\quad
     \pmb{s}_3 =[3\ 4].
   $$
   Suppose that one error occurred in the information part and another in the parity-check part.
Multiplying the syndrome on the left by $I\otimes B^{-1}$, we obtain:


   \begin{equation}\label{eq:ex:doserros}
   \left[
\begin{array}{ccc}
    B^{-1} & 0 &0 \\
     0&B^{-1}&0\\
     0&0&B^{-1}
\end{array}
   \right]\pmb{s}^t=
\begin{bmatrix}
1 & 12 & 0 & 0 & 0 & 0\\
0 & 1  & 0 & 0 & 0 & 0\\
0 & 0  & 1 & 12 & 0 & 0\\
0 & 0  & 0 & 1  & 0 & 0\\
0 & 0  & 0 & 0  & 1 & 12\\
0 & 0  & 0 & 0  & 0 & 1
\end{bmatrix}
\left[
\begin{array}{c}
 9\\
   9\\
   3\\
   3\\
   3\\
   4
   \end{array}
   \right]=
   \begin{bmatrix}
           0\\
    9\\
    0\\
    3\\
   12\\
    4\\
   \end{bmatrix}.
   \end{equation}
   Now, consider the elements in first column of $A$, $A_1=[6\ 4\ 3]^t$, and construct 
$$   diag(6^{-1},4^{-1},3^{-1})=diag(11,10,9).
   $$ 
   Then, multiply the vector \eqref{eq:ex:doserros} by this diagonal matrix: 
\begin{equation}  \label{eq:ex:doserrosb}
(diag(11,10,9)\otimes I_2) \begin{bmatrix}
           0\\
    9\\
    0\\
    3\\
   12\\
    4\\
   \end{bmatrix}=
\begin{bmatrix}
    11 & 0  & 0  & 0  & 0  & 0\\
0  & 11 & 0  & 0  & 0  & 0\\
0  & 0  & 10 & 0  & 0  & 0\\
0  & 0  & 0  & 10 & 0  & 0\\
0  & 0  & 0  & 0  & 9  & 0\\
0  & 0  & 0  & 0  & 0  & 9
\end{bmatrix}
   \begin{bmatrix}
           0\\
    9\\
    0\\
    3\\
   12\\
    4\\
   \end{bmatrix}=
      \begin{bmatrix}
    11\cdot 0\\
    11\cdot 9\\
    10\cdot 0\\
    10\cdot 3\\
   9\cdot 12\\
    9\cdot 4\\
   \end{bmatrix}=
      \begin{bmatrix}
     0\\
    8\\
     0\\
   4\\
   4\\
    10\\
   \end{bmatrix}.
\end{equation}
There are no similar  symbols in the   vector \eqref{eq:ex:doserrosb},    therefore, we pass to the second column and do the same. 
Consider now the elements in second column of $A$, $A_2=[9\ 3\ 3]^t$, and construct 
$$   diag(9^{-1},3^{-1},3^{-1})=diag(3,9,9).
   $$ 
Then we mutiply the vector   \eqref{eq:ex:doserros} by the diagonal matrix:
$$
(diag(3,9,9)\otimes I_2) \begin{bmatrix}
           0\\
    9\\
    0\\
    3\\
   12\\
    4\\
   \end{bmatrix}=
\begin{bmatrix}
    3 & 0  & 0  & 0  & 0  & 0\\
0  & 3& 0  & 0  & 0  & 0\\
0  & 0  &9& 0  & 0  & 0\\
0  & 0  & 0  & 9 & 0  & 0\\
0  & 0  & 0  & 0  & 9  & 0\\
0  & 0  & 0  & 0  & 0  & 9
\end{bmatrix}
   \begin{bmatrix}
           0\\
    9\\
    0\\
    3\\
   12\\
    4\\
   \end{bmatrix}=
      \begin{bmatrix}
    3\cdot 0\\
    3\cdot 9\\
    9\cdot 0\\
    9\cdot 3\\
   9\cdot 12\\
    9\cdot 4\\
   \end{bmatrix}=
      \begin{bmatrix}
     \textcolor{red}{0}\\
    \textcolor{red}{1}\\
     \textcolor{red}{0}\\
   \textcolor{red}{1}\\
   4\\
    10\\
   \end{bmatrix}.
$$
Notice that the first two symbols are equal, therefore that is the value of the error in the information part.
Since we are currently working with the second column, the  error is in position two of the information symbols:
$$ \pmb{e}_2=[0\ 1]. $$ 
Since the third syndrome symbol is different, then we know there is another error in the third parity symbol, that is, the sixth symbol of the received word.
Since we know that 
$$\pmb{e}_2^t+a_{22}^{-1}B^{-1}\pmb{e}_6^t=
\left[
\begin{array}{c}
     4 \\
     10 
\end{array}
\right]
,$$
then we can compute the error:
$$\begin{bmatrix}
     0  \\
     1 
\end{bmatrix}+9B^{-1}\pmb{e}_6^t=
\begin{bmatrix}
     4  \\
     10 
\end{bmatrix}
 \Rightarrow \pmb{e}_6^t=3B\left(\begin{bmatrix}
     4\\10\\
 \end{bmatrix}
 -
 \begin{bmatrix}
     0\\1
 \end{bmatrix}\right)
=
\begin{bmatrix}
    \textcolor{blue}{0}\\
        \textcolor{blue}{1}
\end{bmatrix}.
 $$
The correct codeword is:
\begin{align*}
     \pmb{c} =& \pmb{v}- \pmb{e}\\
     =&
[
\begin{array}{cc|cc|cc|cc|cc|cc}
1 &   0&    1&    1&    0&    0&   11&    0&    6&    0&    7&    1]
\end{array}-
[
\begin{array}{cc|cc|cc|cc|cc|cc}
0 &   0&    \textcolor{red}{0}&    \textcolor{red}{1}&    0&    0&   0&    0&    0&    0&        \textcolor{blue}{0}&        \textcolor{blue}{1}]
\end{array}\\
=&
[
\begin{array}{cc|cc|cc|cc|cc|cc}
1 &   0&    1&    0&    0&    0&   11&    0&    6&    0&    7&     0]
\end{array}. 
\end{align*}

 \end{example}

 Notice that, again, if we consider $B_1=B_2=\cdots=B_k=B$, the complexity of the algorithm is reduced.

 \subsection{Two errors error   in the information symbols}

Consider two symbol errors in the information part, located at distinct
positions $0\leq j_1\leq j_2 \leq k$, with
$\pmb{e}_{j_1}\neq\pmb{0}$ and $\pmb{e}_{j_2}\neq\pmb{0}$. The error vector is
given by
$$
\pmb{e}
=
[
\underbrace{
\pmb{0}\ \cdots\ \pmb{e}_{j_1}\ \cdots\
\pmb{e}_{j_2}\ \cdots\ \pmb{0}
}_{k\text{ symbols}}
\mid
\underbrace{
\pmb{0}\ \cdots\ \pmb{0}
}_{n-k\text{ symbols}}
].
$$

The corresponding syndrome is
\begin{align}
\pmb{s}^t
&=H\pmb{e}^t \nonumber\\
&=
\left[
\begin{array}{cccc|c}
a_{11}B_1 & a_{12}B_2 & \ldots & a_{1k}B_k & I_{b(n-k)}\\
a_{21}B_1 & a_{22}B_2 & \ldots & a_{2k}B_k & \\
\vdots & \vdots & & \vdots & \\
a_{n-k,1}B_1 & a_{n-k,2}B_2 & \ldots & a_{n-k,k}B_k &
\end{array}
\right]
\pmb{e}^t \nonumber\\
&=
\left[
\begin{array}{c}
a_{1j_1}B_{j_1}\pmb{e}_{j_1}^t
+a_{1j_2}B_{j_2}\pmb{e}_{j_2}^t\\
a_{2j_1}B_{j_1}\pmb{e}_{j_1}^t
+a_{2j_2}B_{j_2}\pmb{e}_{j_2}^t\\
\vdots\\
a_{n-k,j_1}B_{j_1}\pmb{e}_{j_1}^t
+a_{n-k,j_2}B_{j_2}\pmb{e}_{j_2}^t
\end{array}
\right].
\label{eq:2erros:1}
\end{align}

To obtain a simpler decoding procedure, we consider the case
$
B_1=B_2=\cdots=B_k=B.
$
Multiplying \eqref{eq:2erros:1} on the left by
$I_{n-k}\otimes B^{-1}$, we obtain
\begin{equation}\label{eq:sys}
I_{n-k}\otimes B^{-1}\pmb{s}^t=
\left[
\begin{array}{c}
a_{1j_1}\pmb{e}_{j_1}^t+a_{1j_2}\pmb{e}_{j_2}^t\\
a_{2j_1}\pmb{e}_{j_1}^t+a_{2j_2}\pmb{e}_{j_2}^t\\
\vdots\\
a_{n-k,j_1}\pmb{e}_{j_1}^t+a_{n-k,j_2}\pmb{e}_{j_2}^t
\end{array}
\right]
=
\left[
\begin{array}{cc}
a_{1j_1}I_b & a_{1j_2}I_b\\
a_{2j_1}I_b & a_{2j_2}I_b\\
\vdots & \vdots\\
a_{n-k,j_1}I_b & a_{n-k,j_2}I_b
\end{array}
\right]
\left[
\begin{array}{c}
\pmb{e}_{j_1}\\
\pmb{e}_{j_2}
\end{array}
\right]=
(A^*\otimes I_b)\left[
\begin{array}{c}
\pmb{e}_{j_1}\\
\pmb{e}_{j_2}
\end{array}
\right],
\end{equation}
where $A^*$ is the submatrix of $A$ formed by the columns indexed by $j_1$ and $j_2$.
The coefficient matrix $A^*\otimes I_b$ in \eqref{eq:sys} has dimensions
$(n-k)b\times 2b$.


Notice that, since $A$ is superregular, the matrix
$A^*\otimes I_b$ has full column rank and therefore admits a left inverse.
Since the system \eqref{eq:sys} has a unique solution, it is sufficient to
consider any $2b$ linearly independent equations. In particular, we can use
the first $2b$ equations, since the corresponding $2\times2$ submatrix of
$A^*$ is non-singular by the superregularity of $A$.

 \begin{example}\label{ex:2erros} Consider the same code as in Example \ref{exemploF13}.   
Assume the same codeword
$$
 \pmb{c}=[
 \begin{array}{cc|cc|cc|cc|cc|cc}
1 &   0&    1&    0&    0&    0&   11&    0&    6&    0&    7&     0]
\end{array}
$$
but now with the received word
$$ \pmb{v}=[
\begin{array}{cc|cc|cc|cc|cc|cc}
1 &   0&    \mathbf{1}&    \mathbf{1}&    \mathbf{2}&    \mathbf{3}&   11&    0&    6&    0&    7&    0].
\end{array}$$
The syndrome is computed by:
$$
\pmb{s}^t=
H \pmb{v}^t=
\left[
\begin{array}{c}
  0\\
   1\\
   8\\
   6\\
   9\\
   4\\
   \end{array}
   \right],
   $$
   that is,
   $$
   \pmb{s}_1=[0\ 1],\ 
    \pmb{s}_2=[ 8\ 6] \quad \text{and}\quad
     \pmb{s}_3=[9\ 4].
   $$
   Suppose both errors occurred in the information part.  
   Multiplying the syndrome   by   $I_3\otimes B^{-1}$:
   $$
   \left[
\begin{array}{ccc}
    B^{-1} &  O &O \\
    O & B^{-1} &  O\\
    O&O& B^{-1}  \\
\end{array}
   \right]\pmb{s}^t=
    \left[
\begin{array}{cccccc}
   1 & 12 &0 &0 &0 &0\\
   0 & 1  &0 &0 &0 &0\\
   0 &0 &  1 & 12 &0 &0\\
 0 &0&   0 & 1  &0 &0  \\
  0 &0 &  0 &0 &  1 & 12  \\
 0 &0 &0 &0&   0 & 1    \\
\end{array}
   \right]
   \left[
\begin{array}{c}
  0\\
   1\\
   8\\
   6\\
   9\\
   4\\
   \end{array}
   \right]=
   \left[
\begin{array}{c}
   12\\
    1\\
    2\\
    6\\
    5\\
    4\\
\end{array}
   \right]
   =\left[
\begin{array}{c}
   a_{1j_1}\pmb{e_{j_1}}^t +  a_{1j_2}\pmb{e_{j_2}}^t \\
      a_{2j_1}\pmb{e_{j_1}}^t +  a_{2j_2}\pmb{e_{j_2}}^t \\
         a_{3j_1}\pmb{e_{j_1}}^t +  a_{3j_2}\pmb{e_{j_2}}^t \\
\end{array}
   \right].  $$
   The previous matricial system can also be seen as:
   $$
   \left[
\begin{array}{c}
   12\\
    1\\
    2\\
    6\\
    5\\
    4\\
\end{array}
   \right]=
   \begin{bmatrix}
       a_{1j_1}& a_{1j_2}\\
         a_{2j_1}& a_{2j_2}\\
           a_{3j_1}& a_{3j_2}\\
   \end{bmatrix}
   \begin{bmatrix}
       \pmb{e}_{j_1}\\
        \pmb{e}_{j_1}\\
   \end{bmatrix}=
   \left[
\begin{array}{ccccc}
  a_{1j_1} &     0  &  a_{1j_2} &  0      \\
  0&  a_{1j_1} &     0  &  a_{1j_2}        \\
     a_{2j_1} &   0     &  a_{2j_2}  &0\\
      0 &   a_{2j_1}      &  0  &a_{2j_2} \\
     a_{3j_1}  &   0               & a_{3j_2}&0\\
          0  &   a_{3j_1}                & 0& a_{3j_2}
\end{array}
   \right] 
      \left[
\begin{array}{c}
e_{j_1}^1\\
e_{j_1}^2\\
e_{j_2}^1\\
e_{j_2}^2\\
\end{array}
   \right].   $$
   Testing the first two columns of $A$, that is, $j_1=1$ and $j_2=2$:
      $$
   \left[
\begin{array}{c}
   12\\
    1\\
    2\\
    6\\
    5\\
    4\\
\end{array}
   \right]=
   \left[
\begin{array}{cccc}
  6& 0 &9& 0\\
  0& 6& 0 &9\\
  4 &0 &3& 0\\
  0& 4 &0 &3\\
  3 &0& 3& 0\\
  0 &3& 0& 3\\
\end{array}
   \right] 
      \left[
\begin{array}{c}
e_{j_1}^1\\
e_{j_1}^2\\
e_{j_2}^1\\
e_{j_2}^2\\
\end{array}
   \right].   $$
We multiply by the left inverse of the matrix in the system:
$$
     \left[
\begin{array}{c}
e_{j_1}^1\\
e_{j_1}^2\\
e_{j_2}^1\\
e_{j_2}^2\\
\end{array}
   \right]  
=
\left[
\begin{array}{cccccc}
2 & 0 & 7 & 0 & 0 & 0 \\
0 & 2 & 0 & 7 & 0 & 0 \\
6 & 0 & 4 & 0 & 0 & 0 \\
0 & 6 & 0 & 4 & 0 & 0
\end{array}
\right]
 \left[
\begin{array}{c}
   12\\
    1\\
    2\\
    6\\
    5\\
    4\\
\end{array}
   \right]=
 \left[
\begin{array}{c}
     12\\
    5\\
    2\\
    4\\
    \end{array}
   \right].
$$
Therefore
$$
\pmb{e}_{j_1}=[12\ 5]
\quad \mbox{and} \quad \pmb{e}_{j_2}=[2\ 4].
$$
Thus,
\begin{align*}
     \pmb{c}&=[
\begin{array}{cc|cc|cc|cc|cc|cc}
1 &   0&    1&   1&    2&    3&   11&    0&    6&    0&    7&    0 
\end{array}]-[
\begin{array}{cc|cc|cc|cc|cc|cc}
  \mathbf{12} &    \mathbf{5}&    \mathbf{1}&    \mathbf{2}&    0&    0&   0&    0&    0&    0&    0&    0
\end{array}]\\
&=[
\begin{array}{cc|cc|cc|cc|cc|cc}
  \mathbf{2} &    \mathbf{8}&    \mathbf{0}&    \mathbf{12}&    2&    3&   11&    0&    6&    0&    7&    0
\end{array}]
\end{align*}
 However, this is not a codeword, since $H \pmb{c}^t\not = \pmb{0}$, therefore, the errors are not in those positions. Testing the second and the third columns of $A$, that is, $j_1=2$ and $j_2=3$:
      $$
   \left[
\begin{array}{c}
   12\\
    1\\
    2\\
    6\\
    5\\
    4\\
\end{array}
   \right]=
   \left[
\begin{array}{cccc}
 9 & 0 & 6& 0  \\
  0 & 9 & 0& 6  \\
 3& 0& 1& 0\\
  0& 3& 0& 1\\
 3 &0& 9&  0 \\
  0 &3& 0&  9 \\
\end{array}
   \right] 
      \left[
\begin{array}{c}
e_{j_1}^1\\
e_{j_1}^2\\
e_{j_2}^1\\
e_{j_2}^2\\
\end{array}
   \right].   $$
  We muliply by the left inverse:
$$
     \left[
\begin{array}{c}
e_{j_1}^1\\
e_{j_1}^2\\
e_{j_2}^1\\
e_{j_2}^2\\
\end{array}
   \right]  
=
\left[
\begin{array}{cccccc}
10&0&5&0&0&0\\
0&10&0&5&0&0\\
9&0&12&0&0&0\\
0&9&0&12&0&0\\
\end{array}
\right]
 \left[
\begin{array}{c}
   12\\
    1\\
    2\\
    6\\
    5\\
    4\\
\end{array}
   \right]=
 \left[
\begin{array}{c}
0\\
1\\
2\\
3
    \end{array}
   \right].
$$
Therefore
$$
\pmb{e}_{j_1}=[0\ 1]
\quad\text{and} \quad
\pmb{e}_{j_2}=[2\ 3].
$$
Thus,
\begin{align*}
  \pmb{c}&=[
\begin{array}{cc|cc|cc|cc|cc|cc}
1 &   0&    1&    2&    3&    4&   11&    0&    6&    0&    7&    0
\end{array}]-[
\begin{array}{cc|cc|cc|cc|cc|cc}
0 &    0&    \mathbf{0}&    \mathbf{1}&    \mathbf{2}&    \mathbf{3}&   0&    0&    0&    0&    0&    0
\end{array}]\\ 
&=[
\begin{array}{cc|cc|cc|cc|cc|cc}
 1 &    0&    \mathbf{1}&    \mathbf{0}&    \mathbf{0}&    \mathbf{0}&   11&    0&    6&    0&    7&    0
\end{array}].
\end{align*}
which is the correct codeword.
 \end{example}

 \begin{remark}
The left inverse of $M=A^*\otimes I_b$ is not unique in general.
However, a particularly simple left inverse can be obtained by considering
only the first $2b$ equations of \eqref{eq:sys}.

Let
\[
A_2^*=
\begin{bmatrix}
a_{1j_1} & a_{1j_2}\\
a_{2j_1} & a_{2j_2}
\end{bmatrix}.
\]
Since $A$ is superregular, $A_2^*$ is non-singular. Hence, we can express the matrix $M$ as follows
\[
M=
\begin{bmatrix}
A_2^*\otimes I_b\\
\widetilde{A}^*\otimes I_b
\end{bmatrix},
\]
where $\widetilde{A}^*$ consists of the remaining $n-k-2$ rows of $A^*$.
Now, we may define
\[
L=
\left[
(A_2^*)^{-1}\otimes I_b
\ \middle|\
0_{2b\times(n-k-2)b}
\right].
\]
Then
\[
LM
=
\big((A_2^*)^{-1}\otimes I_b\big)
\big(A_2^*\otimes I_b\big)
=
I_2\otimes I_b
=
I_{2b},
\]
so $L$ is a left inverse of $M$.

More explicitly, we have that
$$
\delta=\det(A_2^*)
=
a_{1j_1}a_{2j_2}
-
a_{1j_2}a_{2j_1},
\neq 0$$
and
\[
(A_2^*)^{-1}
=
\delta^{-1}
\begin{bmatrix}
a_{2j_2} & -a_{1j_2}\\
-a_{2j_1} & a_{1j_1}
\end{bmatrix}.
\]
Therefore,
\[
L=
\left[
\begin{array}{cc}
 a_{2j_2}{\delta}^{-1}I_b &
- a_{1j_2}{\delta}^{-1}I_b\\[2mm]
-{a_{2j_1}}{\delta}^{-1}I_b &
{a_{1j_1}}{\delta}^{-1}I_b
\end{array}
\ \middle|\
0_{2b\times(n-k-2)b}
\right].
\]

Thus, for fixed candidate positions $j_1$ and $j_2$, the first $2b$
equations are sufficient to determine the two corresponding error symbols,
while the remaining equations can be used to verify whether the obtained
solution is consistent with the complete syndrome.

For instance, in Example~\ref{ex:2erros}, we had the matrix
\[
M=
\left[
\begin{array}{cccc}
9 & 0 & 6 & 0\\
0 & 9 & 0 & 6\\
3 & 0 & 1 & 0\\
0 & 3 & 0 & 1\\
3 & 0 & 9 & 0\\
0 & 3 & 0 & 9
\end{array}
\right].
\]
In this case, $b=2$. Considering the first $2b=4$ rows of $M$,
the corresponding $2\times 2$ submatrix of $A^*$ is
\[
A_2^*=
\begin{bmatrix}
9 & 6\\
3 & 1
\end{bmatrix}.
\]
Over $\mathbb{F}_{13}$, we have
\[
\delta=\det(A_2^*)=4\neq 0 \Longrightarrow \delta^{-1}=10.
\]
Hence
\[
(A_2^*)^{-1}
=
4^{-1}
\begin{bmatrix}
1 & -6\\
-3 & 9
\end{bmatrix}
=
10
\begin{bmatrix}
1 & 7\\
10 & 9
\end{bmatrix}
=
\begin{bmatrix}
10 & 5\\
9 & 12
\end{bmatrix}.
\]
Therefore, a left inverse of $M$ in $\mathbb{F}_{13}$ is
\[
L=
\left[
(A_2^*)^{-1}\otimes I_2
\ \middle|\
0_{4\times 2}
\right]
=
\left[
\begin{array}{cccc|cc}
10 & 0 & 5 & 0 & 0 & 0\\
0 & 10 & 0 & 5 & 0 & 0\\
9 & 0 & 12 & 0 & 0 & 0\\
0 & 9 & 0 & 12 & 0 & 0
\end{array}
\right].
\]
Indeed, over $\mathbb{F}_{13}$,
\[
LM=
\left[
\begin{array}{cccccc}
10 & 0 & 5 & 0 & 0 & 0\\
0 & 10 & 0 & 5 & 0 & 0\\
9 & 0 & 12 & 0 & 0 & 0\\
0 & 9 & 0 & 12 & 0 & 0
\end{array}
\right]
\left[
\begin{array}{cccc}
9 & 0 & 6 & 0\\
0 & 9 & 0 & 6\\
3 & 0 & 1 & 0\\
0 & 3 & 0 & 1\\
3 & 0 & 9 & 0\\
0 & 3 & 0 & 9
\end{array}
\right]
=
I_4.
\]
Thus, the first four equations are sufficient to determine the two
error symbols for the candidate positions $j_1=2$ and $j_2=3$,
whereas the last two equations can be used to verify that the obtained
solution is consistent with the complete syndrome.
\end{remark}

\section{Example with a Vandermonde matrix}

Let $\alpha\in\mathbb{F}_q$ and consider the Vandermonde matrix
\[
A=
\left[
\begin{array}{cccc}
\alpha & \alpha^2 & \ldots & \alpha^k\\
\alpha^2 & \alpha^4 & \ldots & \alpha^{2k}\\
\vdots & \vdots & & \vdots\\
\alpha^{n-k} & \alpha^{2(n-k)} & \ldots &
\alpha^{k(n-k)}
\end{array}
\right].
\]
We assume that $\alpha$ is chosen so that $A$ is superregular. Let
$B\in\mathsf{Mat}_{b\times b}(\mathbb{F}_q)$ be a non-singular matrix.
Then
\[
H=
\left[
\begin{array}{cccc|c}
\alpha B & \alpha^2B & \ldots & \alpha^kB &
\multirow{4}{*}{$I_{b(n-k)}$}\\
\alpha^2B & \alpha^4B & \ldots & \alpha^{2k}B &\\
\vdots & \vdots & & \vdots &\\
\alpha^{n-k}B & \alpha^{2(n-k)}B & \ldots &
\alpha^{k(n-k)}B &
\end{array}
\right]
\]
is the parity-check matrix of an MDS $\mathbb{F}_q$-linear code with
parameters $[n,k,n-k+1]$ over $\mathbb{F}_q^b$.

Assume that two symbol errors occur at positions $j_1$ and $j_2$ in the
information part. The corresponding error vector is
\[
\pmb{e}
=
[
\underbrace{
\pmb{0}\ \cdots\ \pmb{e}_{j_1}\ \cdots\
\pmb{e}_{j_2}\ \cdots\ \pmb{0}
}_{k\text{ symbols}}
\mid
\underbrace{
\pmb{0}\ \cdots\ \pmb{0}
}_{n-k\text{ symbols}}
].
\]
The syndrome is therefore
\begin{align}
\pmb{s}^t
=
H\pmb{e}^t \nonumber=
\left[
\begin{array}{c}
\alpha^{j_1}B\pmb{e}_{j_1}^t+
\alpha^{j_2}B\pmb{e}_{j_2}^t\\
\alpha^{2j_1}B\pmb{e}_{j_1}^t+
\alpha^{2j_2}B\pmb{e}_{j_2}^t\\
\vdots\\
\alpha^{(n-k)j_1}B\pmb{e}_{j_1}^t+
\alpha^{(n-k)j_2}B\pmb{e}_{j_2}^t
\end{array}
\right].
\label{eq:vand-syndrome}
\end{align}

Multiplying the syndrome by $I_{n-k}\otimes B^{-1}$ gives
\[
(I_{n-k}\otimes B^{-1})\pmb{s}^t
=
\left[
\begin{array}{c}
\alpha^{j_1}\pmb{e}_{j_1}^t+
\alpha^{j_2}\pmb{e}_{j_2}^t\\
\alpha^{2j_1}\pmb{e}_{j_1}^t+
\alpha^{2j_2}\pmb{e}_{j_2}^t\\
\vdots\\
\alpha^{(n-k)j_1}\pmb{e}_{j_1}^t+
\alpha^{(n-k)j_2}\pmb{e}_{j_2}^t
\end{array}
\right].
\]
Let
$
\beta=\alpha^{j_1}
$
and 
$
\gamma=\alpha^{j_2}.
$
Then
\begin{equation}
\label{eq:vand-system}
(I_{n-k}\otimes B^{-1})\pmb{s}^t
=
\left[
\begin{array}{cc}
\beta I_b & \gamma I_b\\
\beta^2 I_b & \gamma^2 I_b\\
\vdots & \vdots\\
\beta^{n-k}I_b & \gamma^{n-k}I_b
\end{array}
\right]
\left[
\begin{array}{c}
\pmb{e}_{j_1}^t\\
\pmb{e}_{j_2}^t
\end{array}
\right].
\end{equation}

For fixed candidate positions $j_1$ and $j_2$, consider the first $2b$
equations of \eqref{eq:vand-system}. Their coefficient matrix is $C\otimes I_b$, where
\[
C=
\begin{bmatrix}
\beta & \gamma\\
\beta^2 & \gamma^2
\end{bmatrix}.
\]
Since $A$ is superregular,
$
\det(C)=\beta\gamma^2-\beta^2\gamma
\neq 0.
$
Consequently,
\[
C^{-1}
=
\left(\beta\gamma^2-\beta^2\gamma\right)^{-1}
\begin{bmatrix}
\gamma^2 & -\gamma\\
-\beta^2 & \beta
\end{bmatrix},
\]
and a left inverse of the coefficient matrix in
\eqref{eq:vand-system} is
\[
L=
\left[
C^{-1}\otimes I_b
\ \middle|\
0_{2b\times(n-k-2)b}
\right].
\]
Thus, for each pair of candidate positions $j_1$ and $j_2$, the first
$2b$ equations determine the corresponding error symbols, while the
remaining equations can be used to verify whether the obtained solution
is consistent with the complete syndrome.

Moreover, since $\beta=\alpha^{j_1}$ and $\gamma=\alpha^{j_2}$, the determinant of $C$ can be expressed directly as a power of the primitive element $\alpha$. Indeed,
\[
\det(C)
=
\alpha^{j_1+2j_2}-\alpha^{2j_1+j_2}
=
\alpha^{j_1+2j_2}\left(1-\alpha^{j_1-j_2}\right).
\]
Therefore, since $j_1\not = j_2$, the factor $1-\alpha^{j_1-j_2}$ is nonzero and may itself be written as a power of $\alpha$ by means of    Zech logarithms~\cite{Huber1990}. If $q$ is odd and
\[
Z_{\alpha}(k)=\log_{\alpha}(1+\alpha^k),
\]
then, since $-1=\alpha^{(q-1)/2}$,
\[
\det(C)
=
\alpha^{\,j_1+2j_2+
Z_{\alpha}\!\left(j_1-j_2+\frac{q-1}{2}\right)}.
\]
In characteristic two, this reduces to
\[
\det(C)
=
\alpha^{\,j_1+2j_2+Z_{\alpha}(j_1-j_2)}.
\]
Thus, the inverse factor appearing in $C^{-1}$ can be computed entirely at the exponent level, which may simplify its implementation when a discrete-logarithm or Zech-logarithm table for $\mathbb{F}_q$ is available.

The complete decoding procedure is presented in
Algorithm~\ref{alg:vandermonde} in Appendix~A.
\begin{example}
Consider $\mathbb{F}_8=\mathbb{F}_2(\alpha)$, where $\alpha$ is a root
of
$
p(x)=1+x+x^3.
$
In particular, $\alpha^3=\alpha+1$ and $\alpha^7=1$. Consider the matrix
\[
A=
\left[
\begin{array}{ccc}
\alpha & \alpha^2 & \alpha^3\\
\alpha^2 & \alpha^4 & \alpha^6\\
\alpha^3 & \alpha^6 & \alpha^9
\end{array}
\right]
=
\left[
\begin{array}{ccc}
\alpha & \alpha^2 & \alpha^3\\
\alpha^2 & \alpha^4 & \alpha^6\\
\alpha^3 & \alpha^6 & \alpha^2
\end{array}
\right],
\]
which is superregular over $\mathbb{F}_8$ and let
\[
B=
\left[
\begin{array}{cc}
1&1\\
0&1
\end{array}
\right].
\]
The matrix $B$ is non-singular and, in fact, $B^{-1}=B$. Hence
\[
H=
\left[
\begin{array}{ccc|c}
\alpha B & \alpha^2B & \alpha^3B &
\multirow{3}{*}{$I_6$}\\
\alpha^2B & \alpha^4B & \alpha^6B &\\
\alpha^3B & \alpha^6B & \alpha^2B &
\end{array}
\right]
\]
is the parity-check matrix of an MDS $\mathbb{F}_8$-linear code with
parameters $[6,3,4]$ over $\mathbb{F}_8^2$.

Consider the codeword
\[
\pmb{c}
=
[
1\ 0
\mid
0\ 0
\mid
0\ 0
\mid
\alpha\ 0
\mid
\alpha^2\ 0
\mid
\alpha^3\ 0
].
\]
Suppose that the error vector is
\[
\pmb{e}
=
[
\pmb{\alpha}\ \pmb{0}
\mid
0\ 0
\mid
\pmb{\alpha}\ \pmb{0}
\mid
0\ 0
\mid
0\ 0
\mid
0\ 0
],
\]
so that two symbol errors occur at positions $j_1=1$ and $j_2=3$ of
the information part. The received word is
\[
\pmb{v}
=
\pmb{c}+\pmb{e}
=
[
1+\alpha\ 0
\mid
0\ 0
\mid
\alpha\ 0
\mid
\alpha\ 0
\mid
\alpha^2\ 0
\mid
\alpha^3\ 0
].
\]
Assume we do not know the error vector. 
The corresponding syndrome  of $\pmb{v}$ is
\begin{equation}\label{ex:vand:synd}
    \pmb{s}^t
=
H\pmb{v}^t
=
H\pmb{e}^t
=
\left[
\begin{array}{c}
\alpha\\
0\\
\alpha\\
0\\
\alpha^6\\
0
\end{array}
\right].
\end{equation}
Multiplying by $I_3\otimes B$ we obtain
\[
(I_3\otimes B)\pmb{s}^t
=
\left[
\begin{array}{c}
\alpha\\
0\\
\alpha\\
0\\
\alpha^6\\
0
\end{array}
\right].
\]

Let us first test the candidate positions $j_1=1$ and $j_2=2$. In this
case,
$
\beta=\alpha,
$ and $
\gamma=\alpha^2,
$
and the coefficient matrix is
\[
M_{1,2}
=
\left[
\begin{array}{cccc}
\alpha & 0 & \alpha^2 & 0\\
0 & \alpha & 0 & \alpha^2\\
\alpha^2 & 0 & \alpha^4 & 0\\
0 & \alpha^2 & 0 & \alpha^4\\
\alpha^3 & 0 & \alpha^6 & 0\\
0 & \alpha^3 & 0 & \alpha^6
\end{array}
\right].
\]
The $2\times2$ matrix associated with the first four equations is
\[
C_{1,2}
=
\begin{bmatrix}
\alpha & \alpha^2\\
\alpha^2 & \alpha^4
\end{bmatrix},
\]
whose inverse is
\[
C_{1,2}^{-1}
=
\begin{bmatrix}
\alpha^4 & \alpha^2\\
\alpha^2 & \alpha
\end{bmatrix}.
\]
Therefore,
\[
L_{1,2}
=
\left[
C_{1,2}^{-1}\otimes I_2
\ \middle|\
0_{4\times2}
\right]
=
\left[
\begin{array}{cccc|cc}
\alpha^4&0&\alpha^2&0&0&0\\
0&\alpha^4&0&\alpha^2&0&0\\
\alpha^2&0&\alpha&0&0&0\\
0&\alpha^2&0&\alpha&0&0
\end{array}
\right].
\]
Applying this left inverse to the transformed syndrome gives
\[
\left[
\begin{array}{c}
\pmb{e}_{1}^t\\
\pmb{e}_{2}^t
\end{array}
\right]
=
L_{1,2}(I_3\otimes B^{-1})\pmb{s}^t
=
\left[
\begin{array}{cccc|cc}
\alpha^4&0&\alpha^2&0&0&0\\
0&\alpha^4&0&\alpha^2&0&0\\
\alpha^2&0&\alpha&0&0&0\\
0&\alpha^2&0&\alpha&0&0
\end{array}
\right]
\left[
\begin{array}{c}
\alpha\\
0\\
\alpha\\
0\\
\alpha^6\\
0
\end{array}
\right] =
\left[
\begin{array}{c}
\alpha^2\\
0\\
\alpha^5\\
0
\end{array}
\right].
\]
Hence the first four equations yield the candidate error symbols
\[
\pmb{e}_{1}=[\alpha^2\ 0],
\qquad
\pmb{e}_{2}=[\alpha^5\ 0].
\]
However, 
\[
H\pmb{e}^t=
\left[
\begin{array}{ccc|c}
\alpha B & \alpha^2B & \alpha^3B &
\multirow{3}{*}{$I_6$}\\
\alpha^2B & \alpha^4B & \alpha^6B &\\
\alpha^3B & \alpha^6B & \alpha^2B &
\end{array}
\right]
\left[
\begin{array}{c}
\pmb{e}_1^t\\
\pmb{e}_2^t\\
\pmb{0}\\
\pmb{0}\\
\pmb{0}\\
\pmb{0}\\
\end{array}
\right]
=
\left[
\begin{array}{c}
\alpha B\pmb{e}_1^t + \alpha^2 B\pmb{e}_2^t\\
\alpha^2 B\pmb{e}_1^t + \alpha^4 B\pmb{e}_2^t\\
\alpha^3 B\pmb{e}_1^t + \alpha^6 B\pmb{e}_2^t
\end{array}
\right]
=\left[
\begin{array}{c}
     \alpha  \\
       0 \\
           \alpha  \\
            0  \\
                 1  \\
                   0  \\
\end{array}\right],
\]
which is different from the syndrome vector given in~\eqref{ex:vand:synd}.
Thus, the pair $(1,2)$ is not consistent with the complete syndrome.

Let us now test the candidate positions $j_1=1$ and $j_2=3$. In this
case,
$
\beta=\alpha,
$ and $
\gamma=\alpha^3,
$
and
\[
M_{1,3}
=
\left[
\begin{array}{cccc}
\alpha & 0 & \alpha^3 & 0\\
0 & \alpha & 0 & \alpha^3\\
\alpha^2 & 0 & \alpha^6 & 0\\
0 & \alpha^2 & 0 & \alpha^6\\
\alpha^3 & 0 & \alpha^2 & 0\\
0 & \alpha^3 & 0 & \alpha^2
\end{array}
\right].
\]
The matrix corresponding to the first four equations is
\[
C_{1,3}
=
\begin{bmatrix}
\alpha & \alpha^3\\
\alpha^2 & \alpha^6
\end{bmatrix},
\]
and
\[
C_{1,3}^{-1}
=
\begin{bmatrix}
\alpha^2 & \alpha^6\\
\alpha^5 & \alpha^4
\end{bmatrix}.
\]
Thus,
\[
L_{1,3}
=
\left[
C_{1,3}^{-1}\otimes I_2
\ \middle|\
0_{4\times2}
\right].
\]
Applying $L_{1,3}$ to the transformed syndrome, we obtain
\[
\left[
\begin{array}{c}
\pmb{e}_{1}^t\\
\pmb{e}_{3}^t
\end{array}
\right]
=
\left[
\begin{array}{c}
\alpha\\
0\\
\alpha\\
0
\end{array}
\right],
\]
and therefore
\[
\pmb{e}_{1}=[\alpha\ 0],
\qquad
\pmb{e}_{3}=[\alpha\ 0].
\]
Moreover,  
\[
H\pmb{e}^t=
\left[
\begin{array}{ccc|c}
\alpha B & \alpha^2B & \alpha^3B &
\multirow{3}{*}{$I_6$}\\
\alpha^2B & \alpha^4B & \alpha^6B &\\
\alpha^3B & \alpha^6B & \alpha^2B &
\end{array}
\right]
\left[
\begin{array}{c}
\pmb{e}_1^t\\
\pmb{0}\\
\pmb{e}^t\\
\pmb{0}\\
\pmb{0}\\
\pmb{0}\\
\end{array}
\right]
=
\left[
\begin{array}{c}
\alpha B\pmb{e}_{1}+\alpha^3B\pmb{e}_{3}\\
\alpha^2B\pmb{e}_{1}^t+\alpha^6B\pmb{e}_{3}^t\\
\alpha^3B\pmb{e}_{1}^t+\alpha^2B\pmb{e}_{3}^t
\end{array}
\right]
=
\left[
\begin{array}{c}
\alpha\\
0\\
\alpha\\
0\\
\alpha^6\\
0
\end{array}
\right].
\]
Hence the pair $(1,3)$ is consistent with the complete syndrome given in \eqref{ex:vand:synd} and the
error vector is recovered as
\[
\pmb{e}
=
[
\alpha\ 0
\mid
0\ 0
\mid
\alpha\ 0
\mid
0\ 0
\mid
0\ 0
\mid
0\ 0
].
\]
Consequently,
\[
\pmb{c}
=
\pmb{v}-\pmb{e}
=
[
1\ 0
\mid
0\ 0
\mid
0\ 0
\mid
\alpha\ 0
\mid
\alpha^2\ 0
\mid
\alpha^3\ 0
],
\]
which is the transmitted codeword.

Notice that this example does not imply that every pattern of two
symbol errors can be uniquely corrected by an $[6,3,4]$ code. Rather,
under the assumption that both errors occur in the information part,
it illustrates how the Vandermonde structure can be exploited to
determine the error values for each pair of candidate positions and
how the remaining syndrome equations can be used to test the
consistency of that pair.
\end{example}

\section*{Acknowledgments}
This work was supported by 
supported by 
Conselho Nacional de Desenvolvimento Científico e Tecnológico
(CNPq), Brazil, with process 405842/2023-6.
The first author was supported   by FAPESP with processes 2024/05051-7 and 2024/00923-6.
The second author was supported by project FAPEMIG RED-00133-21.

\bibliography{sn-bibliography}

 \newpage
 
\color{black}

\appendix
\section{Algorithms}

\begin{algorithm}
 \caption{Recovery of the codeword $\pmb{c}$ over the erasure channel}\label{alg:erasure}
 \textbf{Input:}\\  
 Matrices: $B_1,  \ldots, B_k$, A.\\
 Received codeword $\pmb{v}$ completed with zeros.\\
$ \mathcal{I}=\{i_1, i_2, \ldots, i_t\}$:
   Positions of the errors in the parity-check part.\\
 $ \mathcal{J}=\{j_1, j_2, \ldots, j_s\}$: Positions of the errors in the information part.\\

1.  State $M=A\otimes (B_1,\ldots, B_k)$\\
2.  State $H=[M|I_{b(n-k)}]$\\
3.  Compute $\pmb{s}^t=H\pmb{v}^t$\\
4. if $\pmb{s}=\pmb{0}$\\
5. \quad  $\pmb{c}=\pmb{v}$\\
6.  elseif\\
7. \quad Compute $M^*=[A^*  | B^*] \otimes I_b  $\\
8. \quad  Compute $N=(M^*)^{-1}$\\
9. \quad  Compute $[\pmb{f}_1 \ldots \pmb{f}_{n-k}]^t=N\pmb{s}^t$\\
10.\quad State $\pmb{e}=[\pmb{e}_1   \ldots \pmb{e}_n]=\pmb{0}$ \\
11.\quad for $r=1:t$\\
12. \quad \quad Compute $\pmb{e}_{i_r}=B_{i_r}^{-1}\pmb{f}_r$\\
13.\quad endfor\\
14.\quad for $r=1:s$\\
15. \quad \quad State $\pmb{e}_{j_r+k}=\pmb{f}_{t+r}$\\
16.\quad endfor\\
17.\quad Compute $\pmb{c}=\pmb{v}-\pmb{e}$\\
18.  endif\\
\\
  \textbf{Output:} The codeword $\pmb{c}$
\end{algorithm}

\newpage

\begin{algorithm}
\caption{Recovery of the codeword $\pmb{c}$ in the presence of a single error in the information part: distinct matrices $B_1,\ldots,B_k$}
\label{alg:one-info-distinct-B}
\textbf{Input:}\\
Matrices: $B_1,  \ldots, B_k$, A.\\
Nonsingular matrices
$B_1,\ldots,B_k\in\mathsf{Mat}_{b\times b}(\mathbb{F}_q)$.\\
A received word $\pmb{v}$.\\

1. State
$
M=(A\otimes I_b)\operatorname{diag}(B_1,\ldots,B_k).
$\\
2. State
$
H=\left[M\mid I_{b(n-k)}\right].
$\\
3. Compute
$
\pmb{s}^{T}=H\pmb{v}^{T}.
$\\
4. if $\pmb{s}=\pmb{0}$,\\
5. \quad State $\pmb{c}=\pmb{v}$ and stop.\\
6. endif\\
7. for $j=1:k$\\
8. \quad Compute the normalized syndrome
$
(\pmb{u}^{(j)})^{T}
=
(I_{n-k}\otimes B_j^{-1})\pmb{s}^{T}.
$\\
9. \quad Compute the candidate error
$
\widehat{\pmb{e}}_j^{T}
=
a_{1j}^{-1}(\pmb{u}^{(j)}_1)^{T}.$\\
10. \quad if
$
(\pmb{u}^{(j)}_\ell)^{T}
=
a_{\ell j}\widehat{\pmb{e}}_j^{T}
$,
for every $\ell=1,\ldots,n-k$.\\
11. \qquad State
$
\pmb{e}
=
\left[
\pmb{0}\ \cdots\
\widehat{\pmb{e}}_j\
\cdots\ \pmb{0}
\mid
\pmb{0}\ \cdots\ \pmb{0}
\right].
$\\
12. \qquad Compute
$
\pmb{c}=\pmb{v}-\pmb{e}.
$\\
13. \qquad if $H\pmb{c}^{T}=\pmb{0}$\\
14. \quad \qquad return $\pmb{c}$.\\
15. \qquad endif\\
16. \quad endif\\
17. endfor\\


\textbf{Output:}\\
The recovered codeword $\pmb{c}$.

\end{algorithm}
\newpage

\begin{algorithm}
\caption{Recovery of the codeword $\pmb{c}$ in the presence of a single error in the information part: equal matrices 
$B_1=\cdots=B_k=B$}
\label{alg:one-info-equal-B}

\textbf{Input:}\\
Matrices $A$ and $B$.\\
A received word $\pmb{v}$.\\

1. State
$
M=A\otimes B.$\\
2. State $H=[M\mid I_{b(n-k)}].$\\
3. Compute
$
\pmb{s}^{T}=H\pmb{v}^{T}.
$\\
4. if $\pmb{s}=\pmb{0}$,\\
5. \quad State $\pmb{c}=\pmb{v}$ and stop.\\
6. endif\\
7. Compute the normalized syndrome
$
\pmb{u}^{T}
=
(I_{n-k}\otimes B^{-1})\pmb{s}^{T}.
$\\
8. for $j=1:k$\\
9. \quad Compute the candidate error
$
\widehat{\pmb{e}}_j^{T}
=
a_{1j}^{-1}\pmb{u}_1^{T}.
$\\
10. \quad if
$
\pmb{u}_\ell^{T}
=
a_{\ell j}\widehat{\pmb{e}}_j^{T}
$
for every $\ell=1,\ldots,n-k$.\\
11. \qquad State
$
\pmb{e}
=
\left[
\pmb{0}\ \cdots\
\widehat{\pmb{e}}_j\
\cdots\ \pmb{0}
\mid
\pmb{0}\ \cdots\ \pmb{0}
\right].
$\\
12. \qquad Compute
$
\pmb{c}=\pmb{v}-\pmb{e}.
$\\
13. \qquad if $H\pmb{c}^{T}=\pmb{0}$\\
14. \quad \qquad return $\pmb{c}$.\\
15. \qquad endif\\
16. \quad endif\\
17. endfor\\


\textbf{Output:}\\
The recovered codeword $\pmb{c}$.

\end{algorithm}
\newpage

\begin{algorithm}
\caption{Recovery of the codeword $\pmb{c}$ over the symmetric channel: one error in the information part and one error in the parity part}
\label{alg:one-information-one-parity-error}
\textbf{Input:}\\
Matrices $A$ and $B$.\\
Received word $\pmb{v}$.\\

1. State $M=A\otimes B$.\\
2. State $H=[M\mid I_{b(n-k)}]$.\\
3. Compute $\pmb{s}^t=H\pmb{v}^t$.\\
4. if $\pmb{s}=\pmb{0}$\\
5. \quad $\pmb{c}=\pmb{v}$.\\
6. else\\
7. \quad for $i=1:k$\\
8. \qquad Compute $ D_i=
\operatorname{diag}
\left(
a_{1i}^{-1},
a_{2i}^{-1},
\ldots,
a_{(n-k)i}^{-1}
\right).
$\\
9. \qquad  Compute $ \pmb{u}^{t}
=
(D_i\otimes B^{-1})\pmb{s}^t.$\\
10. \qquad if the symbols of $\pmb{u}$ contain a repeated value\\
11. \qquad\quad Identify $j_2$ as the unique position for which $
\pmb{u}_{j_2}\neq\pmb{u}_\ell
 \text{ for all }\ell\neq j_2.
$\\
12. \qquad\quad Set
$
j_1=i.
$\\
13. \qquad\quad Choose any $\ell\neq j_2$ and set
$
\pmb{e}_{j_1}=\pmb{u}_\ell.
$\\
14. \qquad\quad Compute
$
\pmb{e}_{k+j_2}
=
a_{j_2i}B
\left(
\pmb{u}_{j_2}
-
\pmb{u}_\ell
\right).
$\\
15. \qquad\quad State
$
\pmb{e}=
[\pmb{e}_1\ \ldots\ \pmb{e}_n]
=\pmb{0}.
$\\
16. \qquad\quad Set $\pmb{e}_{j_1}$ and $\pmb{e}_{k+j_2}$ to the values computed above.\\
17. \qquad\quad Compute $\pmb{c}=\pmb{v}-\pmb{e}$.\\
18. \qquad\quad break\\
19. \qquad end if\\
20. \quad end for\\
21. end if\\

\textbf{Output:} The codeword $\pmb{c}$.
\end{algorithm}
\newpage

\begin{algorithm}
\caption{Recover $\pmb{c}$ in the Symmetric Channel for a Vandermonde Matrix
(two information errors)}\label{alg:vandermonde}

\textbf{Input:}\\
A superregular Vandermonde matrix $A$.\\
A non-singular matrix $B$.\\
The
received word $\pmb{v}$.\\

1. State $M=A\otimes B$.\\
2. State $H=[M\mid I_{b(n-k)}]$.\\
3. Compute
$
\pmb{s}^t=H\pmb{v}^t.
$\\
4. if $\pmb{s}=\pmb{0}$\\
5. \quad Set $\pmb{c}=\pmb{v}$.\\
6. else\\
7. \quad Compute
$
\pmb{u}^t
=
(I_{n-k}\otimes B^{-1})\pmb{s}^t.
$\\
9. \quad for $j_1=1:k-1$\\
10. \qquad for $j_2=j_1+1:k$\\
11. \qquad\quad Set
$
\beta=\alpha^{j_1}, 
\gamma=\alpha^{j_2}.
$\\
12. \qquad\quad Compute
$
C_{j_1,j_2}
=
\begin{bmatrix}
\beta & \gamma\\
\beta^2 & \gamma^2
\end{bmatrix}.
$\\
13. \qquad\quad Compute
$
C_{j_1,j_2}^{-1}
=
\frac{1}{\beta\gamma(\gamma-\beta)}
\begin{bmatrix}
\gamma^2 & -\gamma\\
-\beta^2 & \beta
\end{bmatrix}.
$\\
14. \qquad\quad Compute the candidate error symbols
$
\left[
\begin{array}{c}
\pmb{e}_{j_1}^t\\
\pmb{e}_{j_2}^t
\end{array}
\right]
=
(C_{j_1,j_2}^{-1}\otimes I_b)
\left[
\begin{array}{c}
\pmb{u}_1^t\\
\pmb{u}_2^t
\end{array}
\right].
$\\
15. \qquad\quad if
$
\pmb{u}_r^t
=
\beta^r\pmb{e}_{j_1}^t
+
\gamma^r\pmb{e}_{j_2}^t
\qquad
\text{for every }r=3,\ldots,n-k
$\\
16. \qquad\qquad State
$\pmb{e}
=
[\pmb{e}_1\ \ldots\ \pmb{e}_n]
=
\pmb{0}.
$\\
17. \qquad\qquad Set $\pmb{e}_{j_1}$ and $\pmb{e}_{j_2}$ to the values
computed above.\\
18. \qquad\qquad Compute
$
\pmb{c}=\pmb{v}-\pmb{e}.
$\\
19. \qquad\qquad break.\\
20. \qquad\quad end if\\
21. \qquad end for\\
22. \quad end for\\
23. end if\\

\textbf{Output:} The codeword $\pmb{c}$.
\end{algorithm}

\end{document}